# Automated Extraction of Energy Systems Information from Remotely Sensed Data: A Review and Analysis

Simiao Ren, Wei Hu, Kyle Bradbury, Dylan Harrison-Atlas, Laura Malaguzzi Valeri, Brian Murray, and Jordan M. Malof

High quality energy systems information is a crucial input to energy systems research, modeling, and decision-making. Unfortunately, precise information about energy systems is often of limited availability, incomplete, or only accessible for a substantial fee or through a non-disclosure agreement. Recently, remotely sensed data (e.g., satellite imagery, aerial photography) have emerged as a potentially rich source of energy systems information. However, the use of these data is frequently challenged by its sheer volume and complexity, precluding manual analysis. Recent breakthroughs in machine learning have enabled automated and rapid extraction of useful information from remotely sensed data, facilitating large-scale acquisition of critical energy system variables. Here we present a systematic review of the literature on this emerging topic, providing an in-depth survey and review of papers published within the past two decades. We first taxonomize the existing literature into ten major areas, spanning the energy value chain. Within each research area, we distill and critically discuss major features that are relevant to energy researchers, including, for example, key challenges regarding the accessibility and reliability of the methods. We then synthesize our findings to identify limitations and trends in the literature as a whole, and discuss opportunities for innovation.

## 1. INTRODUCTION

Access to clean, reliable, and affordable energy is vital to the prosperity and sustainability of modern societies [1]. We are in a time of substantial change for electricity systems. Energy generation capacity must grow to meet growing global demand. Meanwhile, a growing proportion of generation capacity is composed of decentralized and/or variable renewable energy sources that make planning, monitoring and maintenance of energy systems more complex. Effective decision-making and modeling will be crucial to ensure continued energy security for both developed and developing nations. Unfortunately, precise information about energy systems - a vital resource for such modeling and decision-making - is often limited. While in some countries, the information on energy systems is collected through national statistical agencies such as the U.S. Energy Information Administration, comprehensive data sources are not available for most countries and are often incomplete, or only accessible for a substantial fee or through a non-disclosure agreement.

Remotely sensed data (e.g., satellite imagery, aerial photography) have emerged as a potentially rich source of energy systems information [2] that may help close the energy information gap. Remotely sensed data (RSD) are increasingly cheap, abundant, and available in remote geographic locations. These data are also often available at sufficiently high-spatial resolutions to enable the identification and characterization of many types of energy infrastructure including transmission lines/towers [3], power plants [4], and residential solar arrays [5]. Other types of energy system information - beyond just the presence of the infrastructure - can also be inferred from RSD, including potential risks to infrastructure (e.g. vegetation encroachment for power lines

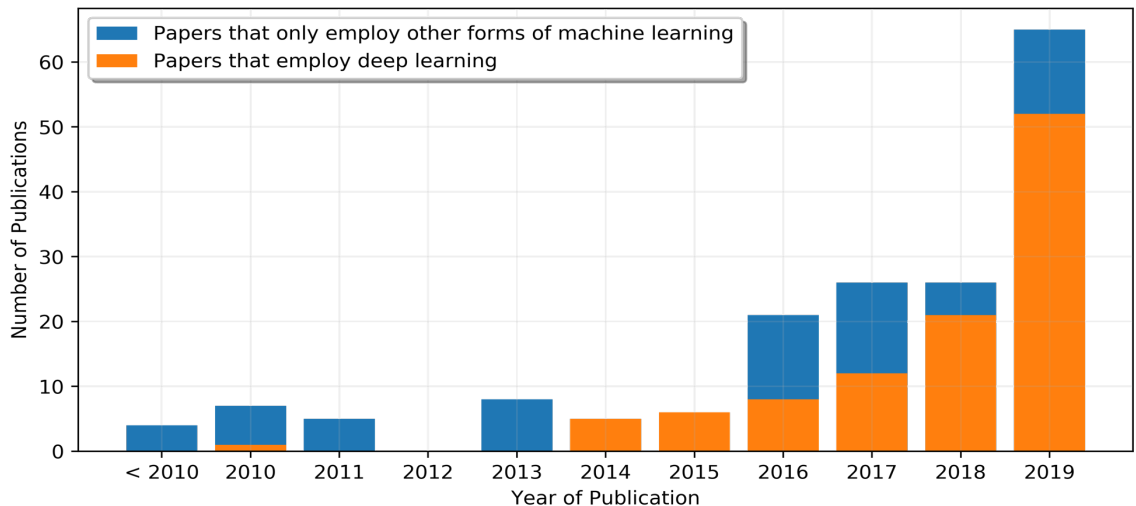

*Figure 1: (1.5 column color) Energy-related publications using Remote Sensing Data (RSD) and Machine Learning (ML) have grown from 2008 to 2019. Publications that rely upon deep learning are shown in blue, illustrating how it has transformed the field.*

[6] ), energy consumption from buildings [7], or energy access [8]. Figure 5 illustrates the variety of energy systems information that has been extracted from RSD.

Although RSD holds tremendous potential, it is far too large for manual analysis. Recent breakthroughs in machine learning (ML) – especially deep neural networks, or "deep learning" [9]–[11] – have made it possible to develop computer algorithms that can automatically and rapidly extract such information from RSD. For example, recent deep-learning-based approaches can map building footprints over an entire U.S. state within just a few hours, and do so using relatively inexpensive hardware often found in personal computers. These approaches are not only fast, they are also often highly accurate, yielding comparable visual recognition accuracy to human analysts on several benchmark tasks [12], [13], [14, p. 10]. Enabled by these technical advances, a large and growing number of ML-based approaches have been proposed to extract energy information from RSD, as shown in Figure 1. In Figure 1, we also stratify the existing literature based upon whether studies employed deep learning or not, illustrating deep learning's increasing popularity and impact on the field.

Collectively, these techniques represent a powerful emerging tool to supply data to inform downstream modeling and decision-making, and thereby close existing information gaps. We refer to this emerging field as REMOTE-ML: **REMOT**ely-sensed **E**nergy systems using **M**achine **L**earning. Despite its high value proposition, energy researchers wishing to utilize REMOTE-ML methods will encounter several challenges: identifying whether energy information of interest can be accurately estimated with them; understanding which ML methods are most suitable for their application; the best choice of RSD for an application and the accessibility of the data; and reliably assessing the accuracy of these methods. Furthermore, even though REMOTE-ML researchers share common technical challenges and scientific goals, it is a nascent field. Consequently, there is limited awareness, conversation, and coordination between REMOTE-ML researchers that have been working within hitherto isolated communities.

In this work, we present a systematic review and critical analysis of this literature, with three primary goals: 1) to increase awareness of, and coordination within, this emerging field; 2) to make the literature more accessible to energy researchers; and 3) to provide a critical analysis and discussion of the challenges and opportunities this emerging field presents. Towards these goals, we conducted a systematic query of Google Scholar using 540 combinations of relevant keywords, from which we identified 185 REMOTE-ML publications published up to March 2020. We then grouped the REMOTE-ML literature into ten highly-represented topical sub-areas of REMOTE-ML, spanning the energy value chain from primary energy resources through end-use consumption (see Section 4 for details). Within each sub-area, we distill and critically discuss major features that are relevant to energy researchers: for example, the accessibility and reliability of the methods, and the key characteristics of energy information that have been extracted using REMOTE-ML techniques. In section 5, we then synthesize our findings to identify challenges and trends in the REMOTE-ML literature as a whole, and discuss opportunities for innovation. For example, in section 5.6, we discuss the security, privacy, and ethical considerations of collecting energy information using REMOTE-ML methods, which is an important shared consideration across the literature.

The remainder of this paper is organized as follows. To make our review more accessible and self-contained, we provide (brief) introductions to essential remote sensing and machine learning concepts, respectively, in Sections 2 and 3. In Section 4, we present our review of the ten REMOTE-ML topical sub-areas. In Section 5, we provide our analysis of REMOTE-ML research as a whole. In Section 6, we present our conclusions and future outlook.

## 2. INTRODUCTION TO REMOTE SENSING DATA SOURCES

Remote sensing deals with the measurement and characterization of objects from a distance. Enabled by increasingly diverse and sophisticated sensors and collection systems, earth imagery has proliferated across both public and proprietary domains. Remote sensors collect data by recording energy emitted or reflected back from the Earth and are often mounted on satellites or airplanes, enabling acquisition of data over broad geographic areas. More recently, data collection has started using drones or unmanned aerial vehicles (UAVs) as well. Sensors generally fall into two classes based on the source of energy used to illuminate the environment. Passive sensors detect energy that is naturally reflected or emitted from the Earth including visible, infrared thermal, and microwave portions of the electromagnetic spectrum. Sunlight typically provides the illumination for passive sensors. In contrast, active sensors such as lidar and radar emit radiation in order to illuminate targeted objects and measure the reflecting radiation. Unique characteristics of the two primary sensor types make them suited to different applications. For example, passive sensors are frequently used to measure properties of the Earth's surface but are ineffective in penetrating cloud cover due to their passive nature. Active sensors are able to penetrate cloud cover and can image in times when sunlight is not present.

Sensor types and vehicles determine the scope and quality of imaging. Understanding these differences is key to identifying which RSD sources best support the extraction of intended energy system information. Important characteristics include the energy content (i.e. part of the electromagnetic spectrum) being captured by the sensor, the spatial resolution of the images, how frequently the sensor revisits the same area (i.e. imaging frequency), geographic coverage, and the sensor vehicle (e.g. satellite or unmanned aerial vehicle, UAV) as summarized in Table 1. A common type of data is RGB imagery, which includes red, green and blue frequency band channels. RGB imagery is relevant for energy applications related to infrastructure detection and characterization, as it enables detection and measurement of objects visible to the human eye, although other sensor modalities such as nighttime light data or infrared, are sometimes also used. The resolution of imagery needed for a given task varies widely, as demonstrated by the examples of energy infrastructure shown at different spatial resolutions in Figure 2. Low resolution imagery (>30m per pixel) enables the identification of areas with some human development, and land cover such as grassland or forests. Medium resolution imagery (10-30m per pixel) can identify some buildings and roads. This level of resolution is used by Landsat 7 and 8 and Sentinel 2, government sponsored satellite programs, which grant free access to the data. High and very high resolution (extending to 0.3-10 meters per pixel or less) allow users to identify the shape of rooftops, the types of vehicles, or the presence of solar PV arrays. These high resolution data are typically provided by a handful of proprietary data providers including Maxar (formerly DigitalGlobe) and Planet for satellite imagery; and AirBus for aerial imagery. As an exception, the National Agriculture Imagery Program (NAIP) is government sponsored, providing free access to high resolution aerial imagery for the United States. Geographic coverage and revisit rate typically varies inversely with resolution: lower resolution data tends to have wider geospatial coverage (e.g. 15-30m/px resolution Landsat data is available globally about every 16 days) while higher resolution data may only collect regional data or if there is global coverage, the revisit rate may be on the order of years, determine the scale and scope of time series analysis and change detection activities. Table 1 summarizes each of these characteristics for data used in REMOTE-ML studies.

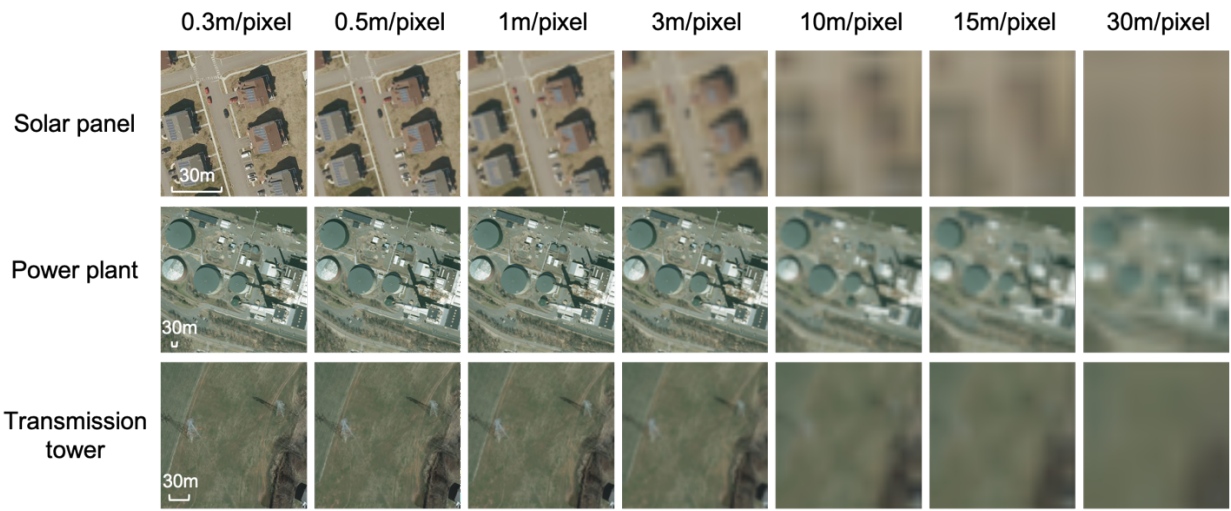

*Figure 2: (2 column color) ) Examples of remotely sensed Red Green Blue (RGB) imagery at varying resolutions showing rooftop solar panels, a power plant, and transmission towers. Very high resolution imagery (< 1 m) is needed to reliably evaluate fine-grain features or to support accurate geometrical characterizations of larger objects. The imagery resolutions correspond to common sensor modalities (prominent data sources listed): 0.3m – highest resolution satellite imagery, proprietary from Worldview 3 or 4; 0.5-1m – NAIP imagery (available openly for the entire U.S.) - otherwise proprietary; 3m - available daily globally from Planet; 10-30m - publicly available globally from Sentinel and Landsat*

| Remote Sensing Data Category | Common Satellites / Data Sources (highest resolution per pixel) [Owner or operator] | Geographic / temporal coverage / revisit time | Energy application examples |
|---|---|---|---|
| Ultra high resolution aerial and drone imagery (<0.3m/px)<br><br>No. of papers: 87 | •UAV (<0.01 m)<br>•Fixed wing aircraft (0.05m-0.1m)<br>•Helicopter (< 0.01 m) | •UAV: flexible<br>•Helicopter: flexible<br>•Plane: flexible | •Power line corridor inspection<br>•Inspection/monitoring of power line components<br>•Damage assessment<br>•Insulator mapping<br>•Solar panel defect assessment<br>•Wind turbine blade defect assessment |
| Very high resolution satellite and aerial imagery (0.3-5m/px)<br><br>Multispectral imagery typically includes RGB imagery, panchromatic, and near infrared bands<br><br>No. of papers: 24 | •Worldview 1-4 (0.31-0.46m) [DigitalGlobe]*<br>•GeoEye-1 (0.41m) [DigitalGlobe]*<br>•Pleiades-1A,1B (0.5m) [Airbus]*<br>•QuickBird (0.61m) [DigitalGlobe]*<br>•IKONOS (0.82m) [DigitalGlobe]*<br>•SkySat-1,2 (0.86m) [Planet]*<br>•NAIP (0.5-2m) [USDA]<br>•SPOT-6,7 (1.5m-5m) [Airbus]* | •Worldview 1-4: global/ available from 2007 to present/ Revisit time: <1 - 2 days<br>•GeoEye-1: global/ available from 2009 to present/ revisit time: 1.7 days<br>•Pleiades-1A, 1B: global / 2012 to present / daily revisit<br>•QuickBird: global / 2001 to 2015 / revisit time: 1-3.5 days<br>•IKONOS: global / 1999 to present / revisit time: 3 days<br>•SkySat-1, 2: global / 2013 to present / revisit time: 4-5 days<br>•NAIP: continental U.S. / available from 2002 to present / Reimage time: 2-3 years<br>•SPOT-6, 7: global / 2012 to present / revisit time: 1-3 days | •Solar PV geo-location and characterization<br>•Transmission & distribution localization<br>•Wind turbine assessment |
| Medium resolution satellite imagery (10-30m/px)<br><br>No. of papers: 14 | •Sentinel 2A,2B (10-20m) [ESA]<br>•Landsat 7 & 8 (15-30m) [NASA / USGS] | •Sentinel: global, 2015 to present, 5-day revisit<br>•Landsat 7&8: global / 1999 to 2021 / 16-day revisit | •Solar Plant detection<br>•Wind resource estimation<br>•Geothermal energy assessment<br>•Coal mine monitoring and identification<br>•Building energy consumption estimation |
| Low resolution satellite imagery (> 30m/px)<br><br>No. of papers: 42 | •MODIS (250m-1km) [NASA]<br>•AVHRR (1.1km) [NOAA]<br>•SSM/I-DMSP (25km) [NOAA] | •MODIS: global / 2013 to present / 2-day revist<br>•AVHRR: global / 1979 to 2021 / daily revisit<br>•SSM/I-DMSP | •Solar potential estimation<br>•Wind potential estimation |
| Nighttime Lights data (NTL) | •VIIRS (375-750m) [NASA / NOAA] | •VIIRS: global, available from 2011 to present, 20-day revisit | •Electricity access assessment |

| No. of papers: 18 | •DMSP-OLS (5km) [NOAA] | •DMSP-OLS: global, available from 1992 to 2013, 12-h revisit | •Electricity reliability estimation<br>•Building energy consumption estimation |
|---|---|---|---|
| Thermal imagery<br><br>No. of papers: 14 | •UAV(<0.01 m)<br>•MTSAT-1R (4km) [JAMI]<br>•SEVIRI [ESA] (3km) | •UAV: flexible<br>•MTSAT-1R: East Asia, Australia, 2005 to present, geostationary<br>•SEVIRI: global, available from 2004, 15 mins revisit | •Solar potential estimation<br>•Solar panel defect detection |
| Radar<br>Synthetic Aperture Radar<br><br>No. of papers: 2 | •Sentinel 1 (5m) [ESA]<br>•Sentinel 3 (300m) [ESA] | •Sentinel 1: global / 2014 to present / revisit time: 12 days<br>•Sentinel 3: global / 2016 to present / revisit time: < 2 days | •Wind potential energy estimation |
| Lidar<br><br>No. of papers: 10 | •Private data collection | •Flexible | •Rooftop analysis for solar placement<br>•Transmission line corridor monitoring |

*Table 1: Overview of remotely sensed data sources for energy applications and their characteristics and illustrative examples of their use for energy system applications. The "No. of papers" listed in the left-most column refer to the number of publications included in our REMOTE-ML literature review that fall into each remote sensing data category.*

## 3. INTRODUCTION TO MACHINE LEARNING FOR REMOTE-ML APPLICATIONS

Machine learning (ML) is a subfield (and sometimes synonym) of artificial intelligence, and generally addresses three types of problems with limited human input: (1) how to make predictions based on data samples (*supervised learning*), (2) how to describe and summarize data (*unsupervised learning*), and (3) how to learn by trial and error (*reinforcement learning*). Although all three of these problems are well-represented in the broader research literature, supervised learning is by far the most common in the REMOTE-ML literature (e.g., it appears in at least 90% of our surveyed literature), and therefore we focus our discussion here on supervised learning. We refer the reader to several textbooks for more thorough treatments of machine learning in genera [15]; as well as supervised learning [16] , unsupervised learning, and reinforcement learning [10].

**Supervised learning.** Supervised learning is essentially the process of approximating some *unknown* function, $y = f(x)$, where $x$ is the explanatory variable, which is input to the function ($f$), and consists of a vector of numerical values. Meanwhile, $y$ is the dependent variable, which is the output of the function, and typically consists of a scalar numerical value, but can also be vector valued. It is assumed that we can evaluate $f$ for specific values of $x$ and thereby obtain the corresponding output $y$ values, but that each of these evaluations is relatively slow and/or expensive. With supervised learning, our strategy is to obtain a small set of $N$ input/output pairs from the function, denoted $D = \{x_i, y_i\}_{i=1}^{N}$. We can then provide these data to a supervised learning algorithm (e.g., linear regression, neural networks), which uses these sample input/output pairs to infer a mathematical expression, denoted $\hat{f}$, that closely approximates $f$. The process of inferring $f$ from $D$ is called "training", and we will discuss it in more detail shortly. The benefit of obtaining $\hat{f}$ is that it is usually substantially faster and cheaper to evaluate than $f$, allowing us to rapidly predict $y$ for novel values of $x$, including values of $x$ that were not in $D$. Note, however, that $\hat{f}$ is only an approximation and will have errors in its predictions. Furthermore, even in problems where $\hat{f}$ is shown to make highly accurate predictions, it will not necessarily have a mathematical form that is similar to the true function $f$.

To illustrate this concept, consider the problem of estimating the energy consumption of a house (i.e., $y$) based upon measurable properties of the home, such as its year of construction, number of floors, square footage, etc. (i.e., these measurements constitute $x$). In general, the relationship between the energy consumption and these variables (i.e., $y = f(x)$)) will be complex and difficult to derive. However, if we can obtain a small but representative sample of daily household energy consumptions, and the corresponding features of the houses (i.e., $D$), then we can use supervised learning algorithms to approximate the true underlying function with a mathematical expression (i.e., $\hat{f}$), which can be used to cheaply and rapidly evaluate the energy consumption of new houses, which have novel values of their properties (often termed "features" in machine learning parlance) that were not present in the training dataset, $D$.

Problems that are amenable to supervised learning arise in a variety of contexts, and supervised learning is now widely used [10]. In the case of REMOTE-ML, the dependent variable is usually some energy-related information, and we want to predict it based upon some explanatory RSD, but the precise relationship between these variables is unknown or uncertain. However, we can usually collect some example data pairs and train a supervised learning model to approximate the relationship. Some examples include the estimation of building energy consumption based upon the area of its footprint in satellite imagery [7], or the solar energy generation potential at a particular geographic location based upon cloud cover density [17].

**Supervised learning for vision tasks.** For many REMOTE-ML applications, supervised learning is applied for computer vision tasks, in which case the explanatory data are imagery. Similar principles apply here as compared to general supervised learning, however, the algorithms are often taxonomized according to the structure of their output, which is unique compared to supervised learning in general. Here we describe three of the most common output structures, or "tasks", in computer vision, and we provide illustrations in Figure 3. In each of these cases, the explanatory input data is always an image, but the structure and size of the dependent output variable is different.

1. **Scene classification**. Classify an image based on its contents. Example question: does this image contain a building? In this case the output is usually a discrete numerical value (e.g., 0 or 1) indicating whether the scene contains a building (a value of one) or not (a value of zero). It is also possible to have more than one discrete assignment categories as well, if there are more possibilities (e.g., building, farm, road, etc.).

2. **Object detection (localization)**. Identify the location of certain types of objects in an image, and often a bounding box indicating the approximate location and size of the objects. Example question: Where are each of the buildings in this image? In this case the output of the algorithm will consist of one vector for each identified building, and each vector will contain numerical values corresponding to the corners of the bounding box.

3. **Image segmentation**. Classify each pixel as being part of one category or another. Example question: is this pixel located over a building? Whereas object detection provides the approximate size of target objects, segmentation can provide the precise shape and size of the target objects.

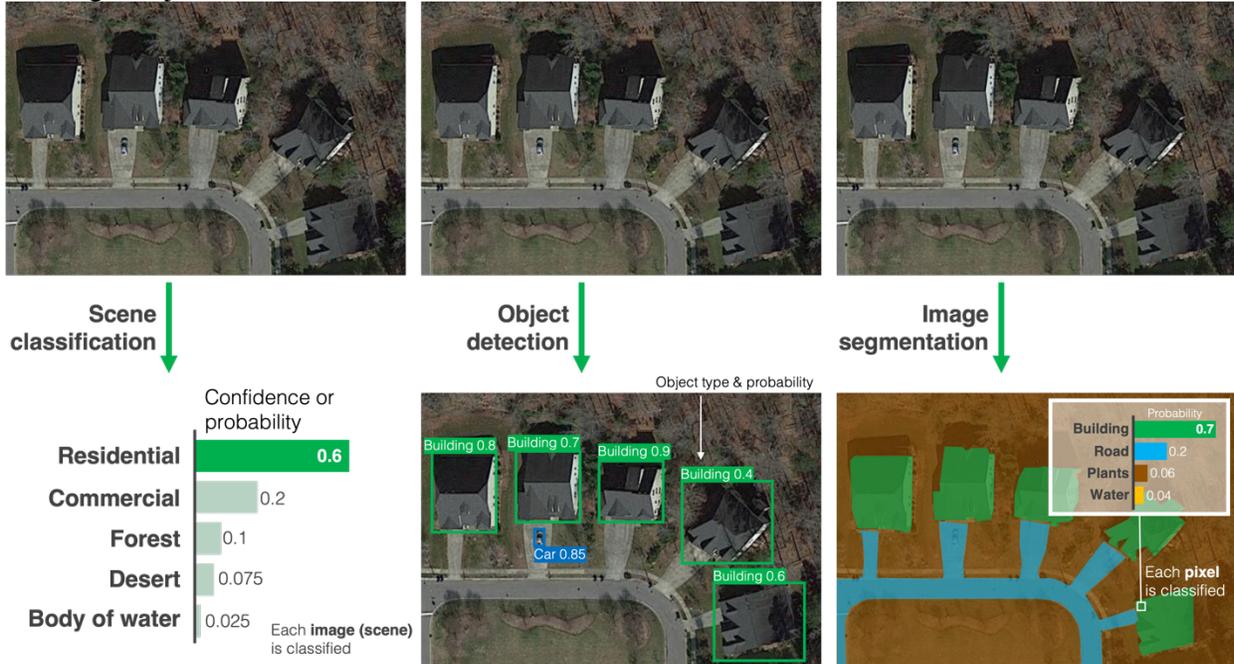

*Figure 3: (2 column color) Overview of the three primary recognition tasks in computer vision for remote sensing data: scene classification, object detection, and image segmentation, with building recognition as a motivating example.*

**Training supervised learning models.** Supervised learning algorithms are composed of a mathematical equation, or model, that has adjustable parameters that control the functional relationship between the input variables and the dependent output variables. By adjusting these parameters, it is possible to approximate many different real-world functions. A simple example is the coefficients of a linear regression model that control the slope and intercept of the regression line. A *training algorithm* is then used to automatically adjust the parameters of the model in an effort to maximize agreement between the model's predictions of the output, $\hat{f}(x)$, and the output values provided with the training data, $f(x)$, over the available training data, $D$. Once the parameters of the model are chosen, the model can be used to make predictions for new values of $x$, including values that were not present in the training data. Supervised learning algorithms utilize different training algorithms, which often have different criteria for "agreement" between the predictions and ground truth, as well as different search procedures for finding the parameters that lead to the best agreement. A discussion of these algorithms is beyond the scope of this paper, and we refer reader to many excellent texts for a more thorough treatment (e.g., [10, p. 20]).

**Evaluating the accuracy of supervised learning algorithms for REMOTE-ML applications.** Once a supervised learning algorithm is trained it can be applied to novel input data, however, there are no general guarantees regarding its prediction accuracy. Furthermore, the prediction accuracy of an algorithm can vary depending upon many factors such as the quantity of training data (more is generally better), the complexity of the function being approximated (lower is generally easier), the learning algorithm that is being utilized, and more. Therefore, it is most common to evaluate the accuracy of a trained algorithm empirically, by evaluating its performance on some testing dataset. A well-known limitation of machine learning models, especially those with many parameters, is their tendency to achieve optimistic prediction accuracy if their performance is evaluated on the same data used for training - a problem often referred to as "overfitting". Therefore, it is crucial to utilize a testing dataset that is *disjoint* from the training dataset. There are many standardized procedures for partitioning available data into disjoint training and testing datasets for this purpose, including random partitioning, or k-fold cross-validation. Another important practice is to utilize a testing dataset that is *representative* of the data that will be encountered by the trained model in its end-use application. For example, in REMOTE-ML applications it is common to apply a trained model to RSD collected over new geographic locations (relative to the training data). Therefore, to obtain an accurate estimate of the model's performance, it should be evaluated on a representative collection of imagery from these locations. This is discussed further in Section 5.1.

One additional consideration when evaluating trained REMOTE-ML algorithms is the performance metric. A variety of different performance metrics have been employed in the REMOTE-ML literature. Some common metrics, as well as their meaning and use are shown in Table 2. Among these common metrics, the most appropriate performance metric will usually depend upon many factors, including the downstream application for the trained algorithm, and the preferences of the user. For example, some of these metrics are only applicable to certain types of problems (e.g., ROC curves, precision, and recall are only applicable to binary classification problems). Another consideration is that each of these metrics accounts for errors somewhat differently, and therefore each metric can yield a different rank-ordering of the same candidate algorithms based upon their testing errors.

| Use | Metric | Description |
|---|---|---|
| Regression | Mean Square Error (MSE) | Mean of the *square* of the difference between the predicted value and true value for all samples |
| | Mean Absolute Error (MAE) | Mean of the *absolute value* of the difference between the predicted value and true value for all samples |
| Classification | Overall Accuracy | Fraction of samples correctly classified (binary or multiclass) |
| | Recall (a.k.a. True Positive Rate; TPR) | Fraction of positive class samples correctly predicted as positive |
| | False Positive Rate (a.k.a. False Detection Rate; FPR) | Fraction of negative class samples incorrectly predicted as positive |
| | Precision | Fraction of positive predictions that are true positives |
| | Receiver Operating Characteristic (ROC) Curve | A plot of TPR versus FPR as a function of the decision threshold, showing all possible modes of performance for a classifier. |
| | Area Under the Curve (AUC) | The area under an ROC curve. Higher values indicate better discriminatory performance with AUC = 1 indicating perfect classification. |
| | Precision Recall (PR) Curve | A plot showing precision versus recall as the decision threshold of the classifier is varied, showing all possible modes of performance. |
| | Average Precision (AP) | The area under a PR Curve. Higher values indicate better performance with AP = 1 indicating perfect classification. |
| | Mean Average Precision (mAP) | For multiclass classification problems, mAP is the average of class-wise binary AP scores, averaged across all classes. |
| Segmentation | Intersection over Union (IoU) (a.k.a. Jaccard Index) | The goal of object detection is to identify a set of pixels that contains an object of interest. Typically, the predictions are in the form of a bounding box. IoU evaluates the quality of these predictions by identifying how much of the area associated with the prediction intersects with the ground truth area and dividing this by the union of the prediction and ground truth area. Higher fractional values indicate better detection performance and the best possible value is IoU = 1. |

*Table 2: Table of common evaluation metrics used for remote sensing applications of machine learning. In the table above, in the case of binary classification metrics, we adopt the convention that the two classes are the "positive" and "negative" class to distinguish between the two.*

**Common supervised learning algorithms.** A large number of supervised learning algorithms have been employed in the REMOTE-ML literature. Some of the most widely-used algorithms include Linear Regression, Decision Trees, Random Forests, (Deep) Neural Networks, and Support Vector Machines. There are also many variants of each of these algorithms. The best choice of algorithm is highly dependent upon the properties of the problem (the complexity and structure of the data), data availability and the needs of the user (e.g., monetary or time constraints).

## 4.   REVIEW OF EXISTING REMOTE-ML WORK

In this section we present a taxonomy of existing REMOTE-ML research, and summarize the work that has been done in each major area of research in our taxonomy. REMOTE-ML research is defined here as addressing energy-relevant questions using ML techniques to analyze RSD from satellite, aerial, or unmanned aerial vehicle (UAV) platforms. This study focuses on applications related to planning, operations and maintenance and evaluation, but does not include methods for analyzing environmental impacts of energy systems, such as the identification of leaks or spills. We searched for relevant RMSLE papers using a systematic query of Google Scholar, which is summarized in Figure 4. We used 540 combinations of keywords from a word bank (e.g., "remote sensing" and "energy" and "machine learning"), and then aggregated the results of each search. From this process we identified 185 REMOTE-ML publications. Additional details for the literature review can be found in Appendix A (e.g., exclusion/inclusion criteria, and search keywords).

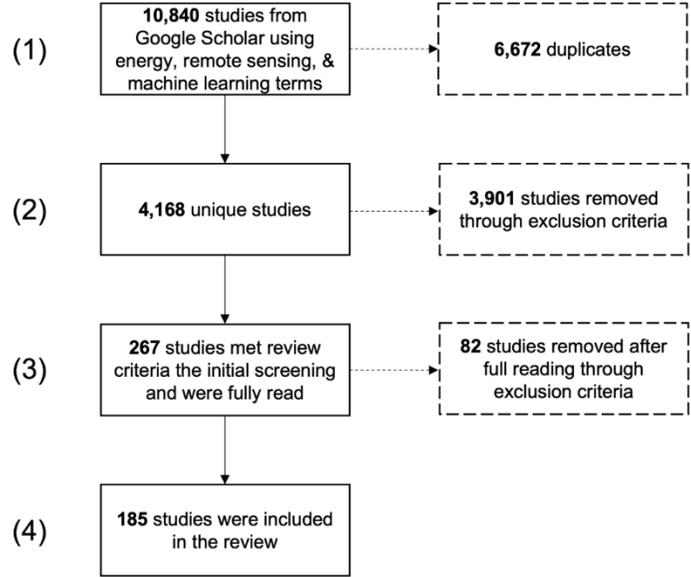

*Figure 4: (2 column color) Flowchart of the literature search and filtering process used to identify the REMOTE-ML publications included in our review. In step 1, a combination of manual search and automated search using keyword pairs (Appendix A for all keyword pairs) was used. In step 2, irrelevant papers were excluded if, after manually reading the paper title and abstract, it was determined that they do not fulfill our inclusion criteria. In step 3, after fully reading all papers, an additional 61 papers were excluded because they did not meet all inclusion criteria.*

In support of our analysis, we reviewed each publication and recorded useful attributes relating to four broad criteria: (i) explanatory RSD; the (ii) the machine learning methods; (iii) the energy information that was extracted; (iv) and other useful characteristics, such as their taxonomic classification (described next). The resulting inventory of publications and their attributes can be accessed or downloaded as a public Google document[1]. Based upon our review of the literature, and aided by our inventory, we then taxonomized the existing literature based primarily upon the energy application described in each paper. For example, two publications would be classified in the same group if they focused on a similar energy application (e.g., solar energy generation potential), even if they utilized different RSD and ML models. This design is intended to help energy researchers navigate the taxonomy (and therefore the literature) most easily. Using this rationale, we identified four major emerging areas of research, and ten major emerging sub-areas of research. The taxonomy and the proportion of papers in each category are presented in Table 3. We wish to emphasize that the ten topical areas identified here were obtained by grouping energy applications found in the *existing published* REMOTE-ML literature, and therefore it does not necessarily include all energy areas. Furthermore, many methodologies from the machine learning and remote sensing literature may plausibly be applied for energy applications and we highlight a few promising ones in sections 5.4 and 5.5, however our review focuses upon those utilized in existing REMOTE-ML publications.

| Level 1 Category | Number of Studies | Level 2 Category | Number of studies |
|---|---|---|---|
| | 61 | Solar Potential Energy Estimation and Forecasting | 34 |

---

| | | | |
|---|---|---|---|
| Primary Energy Resources | | Wind Resource Estimation and Forecasting | 9 |
| | | Other Primary Resources: coal, geothermal, hydroelectric, and oil | 18 |
| Electricity Generation Infrastructure | 32 | Electricity Generation Infrastructure identification and characterization | 16 |
| | | Electricity Generation Infrastructure Risk Monitoring | 16 |
| Electricity Transmission and Distribution Infrastructure | 72 | Transmission and Distribution Line Mapping | 31 |
| | | Transmission infrastructure Mapping Mapping & Monitoring | 12 |
| | | High-voltage Insulator Monitoring | 29 |
| Electricity Consumption | 24 | Building Energy Consumption Estimation | 13 |
| | | Electricity Access and Reliability Estimation | 11 |

*Table 3: Taxonomy of energy applications used to structure our review of REMOTE-ML research. For each of the two levels within the hierarchy, we highlight the number of publications evaluated within that topical area. Note that due to their broad scope, four papers are counted in multiple level II categories.*

In the remainder of this section we review and analyze the literature of each topical sub-area of our taxonomy.

We focus on analyzing each sub-area separately because the literature therein involves relatively similar methodologies and energy applications. Each sub-area review employs the following structure (i.e., sub-headings): *introduction*, *methodological summary*, *energy information*, and *accessibility and reproducibility*. In the *introduction* we explain what types of energy information are being extracted, why this energy information is useful, and the limitations of conventional (non-REMOTE-ML) methods for obtaining it. In the *methodological summary* we explain how the targeted energy information can be inferred using RSD and ML technologies. We then describe the types of RSD that are typically utilized as explanatory data (i.e., input) for the ML models. Next, in the *energy information*, we summarize which energy information has been extracted using these methodologies (e.g., geographic locations, and spatial resolution). In some cases there may be multiple *energy information* headings, if there are fundamentally different types of energy information extracted or estimated by the methods. *Accessibility and reproducibility* describes how easily researchers can replicate reported results in the literature, or utilize the proposed methodologies in new contexts (e.g., novel geographic locations). These sections usually focus on the ease of implementing the processing or ML methods, as well as the accessibility of the RSD upon which the method relies - these are often the two biggest factors influencing accessibility and reproducibility.

## Primary Energy Resources

Solar energy estimation and forecasting

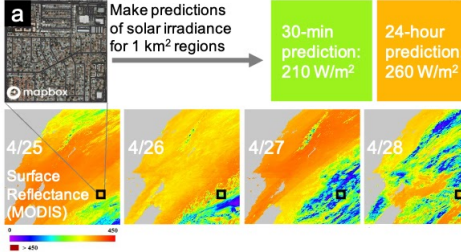

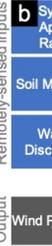 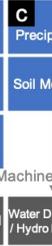 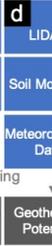 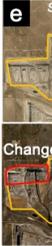 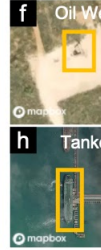

## Generation Infrastructure

Energy generation infrastructure identification

Energy generation infrastructure risk monitoring

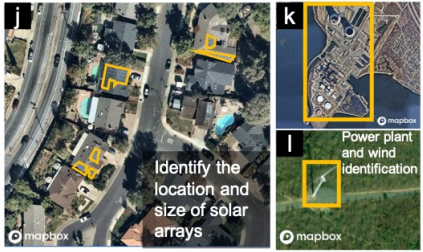

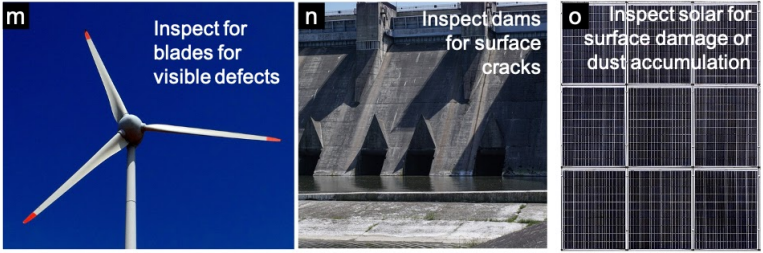

## Transmission Infrastructure

Transmission & distribution line localization

Transmission line corridor risk monitoring

High voltage insulator monitoring

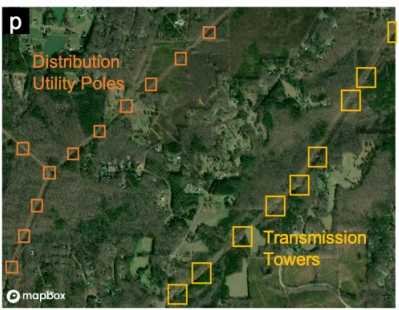 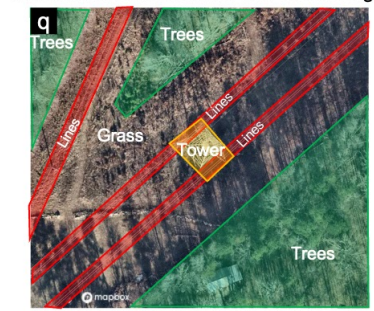 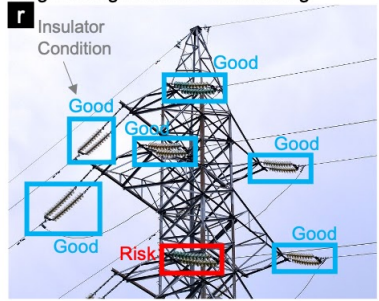

## Electricity Consumption

Building energy consumption

Electricity access and reliability estimation

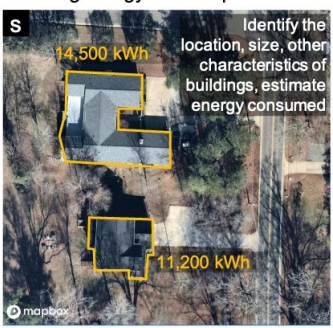 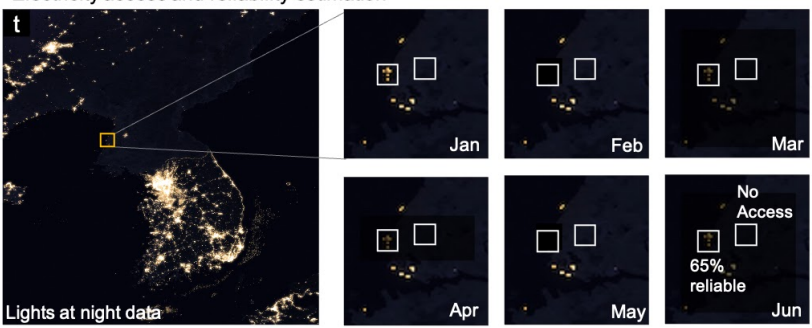

*Figure 5: (2 column color) ) Illustration of the primary REMOTE-ML applications. (a-i) mapping, or characterizing, monitoring, and forecasting primary energy resource including solar, wind, hydro, geothermal, coal, oil, and natural gas, described in Section 4.1; (j-o) mapping, assessing risks, and monitoring energy generation infrastructure such as solar photovoltaics, wind turbines, and hydro dams, described in Section 4.2; (p-r) mapping electricity transmission infrastructure, monitoring transmission corridor risks monitoring and assessing high voltage insulator health, described in Section 4.3; and (s-t) estimating electricity consumption, access and reliability, described in Section 4.4. Sources of satellite imagery used in this figure include: (a, f, g, h, i, j, k, l, p, r) from Mapbox (© Mapbox, © OpenStreetMap, © Maxar), and (e) Planet.*

### 4.1. Primary energy resources

#### 4.1.1. Solar potential energy estimation and forecasting

*Introduction.* In this section we discuss REMOTE-ML techniques to estimate and forecast solar energy generation potential. One important measure of potential solar energy, and the focus of all REMOTE-ML-based methods, is solar irradiance. Solar irradiance is broadly defined as the quantity of solar energy incident at a particular time and place on the surface of the Earth. Solar irradiance estimates usually represent an *average* of solar irradiance that is expected over some length of time, referred to here as the "time interval" of the estimate. Irradiance estimates may be for historic time periods or may be forward-looking. Predictions for future quantities of solar irradiance are termed "forecasting", and the length of time into the future that is captured by a forecast is referred to as its "forecast horizon".

Solar irradiance estimates are useful for a variety of different applications depending upon their temporal scope (historic vs future) as well as their forecast horizon. For example, estimates of historical solar resources may be used to represent the long-term solar potential of a particular location. Consequently, these estimates are often used for solar project planning [18]. Forecasts with horizons ranging from hours to days play a significant role in solar project management like grid balancing, transmission scheduling and clearing of day-ahead electricity market [19]. Short-term forecasts with horizons of a day or less can help system operators balance supply and demand at each moment in time, resulting in more efficient grid operation [20] and an easier integration of variable solar resources into the power grid.

Traditionally, solar irradiance is measured using pyranometers, which are ground-based instruments that measure the quantity of energy received from the sun within a specific portion of the electromagnetic spectrum. These devices are usually located at meteorological stations, which are relatively scarce, thereby limiting the spatial resolution and coverage of irradiance measurements. Motivated by this limitation, researchers have investigated REMOTE-ML approaches as a means to obtain irradiance estimation with greater geographic coverage and resolution (illustrated in Figure 5(a)). For further information on ground-based irradiance estimation and forecasting we refer the reader to related reviews [19], [21].

*Methodological Summary.* Although solar irradiance cannot be directly measured from RSD platforms (e.g., satellites), it can be inferred indirectly because the energy from the sun reaching the outer atmosphere of the earth is approximately constant. As a result, any variations of ground level irradiance must be caused by earth's atmosphere (e.g., sunlight being reflected back, or scattered) – here termed *atmospheric loss*. Therefore, irradiance can be indirectly inferred using a general two-step process: (i) infer the atmospheric loss using remote sensing measurements that are correlated with this loss (e.g., cloud cover); then (ii) subtract the atmospheric loss from the energy reaching the earth. In this way, RSD can be used to indirectly estimate ground-level irradiance, although we note that the inference process can be more complicated than the two-step example described above. Because RSD can be collected at a relatively high spatial resolution (e.g., 1 km$^2$) over very large areas, it is possible to obtain irradiance estimates with similarly high geographic coverage and resolution. By contrast, conventional ground-based estimation methods are limited in their numbers and density globally, providing much lower coverage and resolution.

ML methods are employed in the irradiance estimation process because the precise relationships between RS measurements and atmospheric loss can be complex, or altogether unknown. ML methods can be trained to infer the atmospheric loss, or (more commonly) to directly infer the irradiance, based upon explanatory data that often include a time-series of current and past remotely-sensed measurements and/or ground-based measures of irradiance. Most often, the time-series of measurements consists of optical imagery, (less often) infrared imagery, or land surface temperature measurements. With modern geo-stationary satellites such as the MSG [20] and GOES, these measurements can be obtained at (up to) 15-minute intervals, and 1 km$^2$ or less spatial resolution. ML algorithms are then trained to learn the relationships between these input data and atmospheric loss using paired observational data consisting of remotely-sensed measurements and ground truth irradiance measurements. ML techniques used for this purpose include Support Vector Machines [22]–[32], Random Forest [18], [22], [31]–[33], and Neural network [22], [23], [25], [26], [34]–[39].

Although most of the literature on this topic has focused upon solar irradiance estimation, some work has also been done to estimate other related quantities such as irradiance variability, and rooftop solar energy potential. We therefore divide our energy information summary into two parts: irradiance estimation, and estimation of other quantities.

*Energy information: Irradiance Estimation.* For short term solar irradiation / photovoltaic (PV) output prediction, researchers [27], [30], [32], [40]–[42] have explored methods to forecast solar irradiance with (approximately) hourly time intervals and horizons in Korea, China, and Japan with a spatial resolution of ~1 km. Others [20], [23], [25], [29], [37], [43]–[45] used MSG satellites to make forecasts with hourly time intervals, and forecast horizons ranging from one to six hours, and ~3km resolution in Spain, Greece, and the Netherlands. Similar forecasts have also been explored in North and South America [31], [46] , Africa [31] and Oceania [47]. Researchers have explored medium-term (daily) estimates of irradiance in several parts of Europe such as Spain [34], [48] and France [36] . Using publicly available data from JMA or MSG satellite systems, researchers have also estimated daily irradiance with ~1km resolution in China [17], [49], Turkey [28], [50] and Singapore [39]. In these medium-term studies, the forecast horizon is rarely specified. For longer-term solar irradiance estimation, Australia is the most active region, with multiple methods being compared across multiple cities at a spatial resolution of 1km [18], [26], [33], [38] . MODIS satellite data were also used for estimating monthly solar irradiance in Colombia [22], and Sahin et al. [51], [52] studied monthly irradiance estimation in Turkey with a spatial resolution of 1.1 km.

*Energy information: Other Quantities.* Zagouras et al. [53] characterized the variability of solar irradiance in Greece to support the selection of new ground-based irradiance measuring stations. Another set of research has focused upon identifying rooftops in urban areas that are most suitable (e.g., due to their structural stability, or average daily irradiance) for the installation of solar electricity generation hardware. Methods to estimate rooftop area or potential have been developed and applied to Bangalore, India [54]; Wuhan, China [55] ; and Geneva, Switzerland [24], [56, p.].

*Accessibility and Reproducibility.* Most of the methods proposed in this literature rely upon explanatory data that is publicly available (e.g., MODIS and MSG-2 imagery), overcoming a major obstacle for REMOTE-ML research, and making these methods more accessible. This is a major reason why these methods have been deployed across such a large and diverse number of locations. By contrast, very few publications (<5%) published the code needed to produce the results reported in the publication. This may present a relatively large obstacle for irradiance estimation research because the pre-processing descriptions in publications are sometimes vague, and the accuracy of the proposed methods can be sensitive to changes in the pre-processing especially when the preprocessing step contains feature selection [24]. One additional challenge is access to ground truth measurements because this information is often not centralized, and therefore may require the cooperation of private and/or government organizations or research efforts for curation of these data to provide large-area coverage.

### 4.1.2. Wind resource estimation and forecasting

*Introduction.* In this section we discuss the estimation of wind resources, which are generally quantified in terms of the velocity of the wind at a particular location and hub height over time. The generation of electrical power based upon wind is generally proportional to the cube of the wind's velocity, subject to turbine operating characteristics, making accurate estimation of wind resources of critical importance. Wind estimates are usually given in terms of the *average* wind velocity over some length of time, referred to here as its "time interval". Similar to solar resource estimation, wind resource estimation may involve characterization of historic wind speeds or may be aimed at predicting future conditions (i.e., forecasts).

Characterizations of wind resources are useful for a variety of applications depending upon whether they are historical or forward-looking in nature. Different forecast horizons also have unique applications (e.g., see [57] for a detailed list). For example, multi-year historical assessments describe long-term wind resources at a particular location. Such robust historical wind resource characterizations are often used to plan the installation of wind farms (e.g., in locations that have high monthly or annual wind potential estimates). Alternatively, forward-looking wind forecasts with short time horizons (e.g., hourly or daily) are used by grid operators to balance grids with large wind power penetration [58], [59], for instance, by bringing generators online or offline [57].

Traditionally wind velocity estimates are obtained by anemometers (i.e., a device to measure wind speed and direction) that are located at ground-based meteorological stations\masts, or on buoys located on bodies of water. Wind velocity forecasts are then made by extrapolating forward in time to the desired forecast horizon based upon historical anemometer measurements, or with numerical wind flow models that rely on physical models [60]. Due to the cost of installing and maintaining anemometers [61], however, they are only placed sparsely around the globe, and they only effectively measure the wind velocity in close proximity to the anemometer itself [62]. A more recent measurement approach relies on ground-based SODAR/LiDAR, which can take wind measurements within a vertical column above the measurement device, thereby providing superior coverage (per sensor) compared to anemometers [61], [63], [64]. However, SODAR/LiDAR measurements can only be collected where line-of-site and suitable weather conditions are present, ultimately limiting their geographic coverage and density of measurements. The limitations of these existing estimation mechanisms has motivated the use of REMOTE-ML techniques using airborne *(*especially satellite*)* measurements, which can produce a dense spatial grid of estimates covering large geographic areas, complementing the coverage of ground-based measurements, and improving upon the accuracy of predictions made by some physical models [46].

*Methodological Summary.* Although wind velocity cannot be directly measured from remote sensing platforms (e.g., satellites), it can still be inferred indirectly because many types of remote sensing measurements correlate with wind speed. A variety of such measurements have been utilized in this capacity; however, due to space limitations, we will summarize a few of the most important measures here. One notable example is the use of LiDAR on satellites, which can be used to estimate ground elevation maps (e.g., the AsterDEM [62] Wind velocity is affected by local topography and therefore these elevation maps can contribute to models that estimate local wind velocity. Also, certain types of satellite-based synthetic aperture radar (SAR) imagery (e.g., ENVISAT), can be used in conjunction with geophysical models to characterize wind resources over open bodies of water by measuring the scattering of radar from the wave [65]. This method relies on the fact that stronger winds lead to more wave activity, and therefore greater levels of radar scattering. Similarly, the intensity of microwave imaging (e.g., SSM/I) varies depending upon wave and foam intensity on the surface of the ocean, which depends upon wind velocity. Surface-level estimates of wind speed require vertical extrapolation to different hub heights for wind energy applications. ML methods are employed in the inference process (as well as for vertical extrapolation [66]) because the precise relationships between many of these RSD and wind characteristics are complex, or altogether unknown. ML methods are therefore trained to estimate the wind velocity based upon a time-series of current and past RS measurements. Various ML algorithms have been employed in this role such as Random Forests [67], [68], Neural Networks [69]–[72], Support Vector Regression [73], and k-Nearest Neighbor [73].

*Energy Information.* In the 1990s, seminal work by [69]–[71] mapped wind speed in the Pacific Ocean with a low spatial resolution of >25km and temporal resolution of ~18h with the use of SSM/I sensors. Subsequently, using more recent RS hardware Stefano et al. [67] used solar and meteorology data (e.g., air temperature, relative humidity and atmospheric pressure etc.) from satellite measurements to predict the long-term average wind velocity at 1 km$^2$ spatial resolution in Switzerland. Concurrently, He et al. [73] combined synthetic-aperture radar (SAR) data onboard ENVISAT and ECMWF wind velocity data to provide wind velocity data with high spatial-temporal resolution in Norway. Veronesi et al. [62] used satellite-derived elevation, land cover and solar irradiance explanatory data to characterize average long-term wind speed at 1 km$^2$ resolution in England. The same group further extended this idea [68] to predict long-term average wind velocity around much of the globe. Recently, with the use of high-resolution (5m) satellite SAR imagery from Sentinel-1, Majidi et al. [72] estimated near and off-shore wind velocity for Sardinia, Italy. Apart from wind speed, Harrison-Atlas et al. [74] also investigated wind power capacity density in the U.S. with 2km resolution.

*Accessibility and Reproducibility.* The wind potential forecasting methods surveyed here are quite accessible to other researchers. The vast majority of explanatory data in these methods is freely and publicly available. It is notable that some RSD platforms used in the surveyed papers (e.g., DMSP, ENVISAT) have since been terminated and are therefore no longer accessible. However, newer and higher resolution (ENVISAT has a SAR resolution of 30m while Sentinel-1a has SAR resolution of 5m) RSD systems are currently in operation (e.g., Sentinel).

### 4.1.3. Other primary energy resources: coal, geothermal, water, and oil

*Introduction.* This section reviews applications focused on the characterization of four types of primary electricity generation resources, or closely-related infrastructure: coal, geothermal, hydroelectric, and oil. These studies represent nascent but promising REMOTE-ML use-cases for which there are relatively few publications (17 identified through this review), with most of those (67%) published in the last three years. Given the relatively limited work on each of these topics, we summarize each of them using a single paragraph below. We note that environmental impacts associated with energy systems infrastructure and operations (e.g., classification of land conversion [75], detection of gas leaks [76], oil spills and emissions [77]) are critical topics for which REMOTE-ML methods can play a meaningful role in illuminating the broader sustainability of energy systems, but are not included here for reasons of scope (see APPENDIX A for details).

*Geothermal resource estimation.* These methods attempt to detect the presence of geothermal potential energy (e.g., identifying geothermal anomalies), or to estimate the quantity of geothermal potential energy (e.g., ground temperature). Conventional approaches for estimating these quantities, such as perforation (e.g., drilling a hole to measure temperature), are costly and time-consuming [78]. REMOTE-ML methods leverage known or hypothesized relationships that exist between geothermal potential and RSD to estimate geothermal potential using only the RSD. Ramdhan et al. [79] combined airborne thermal infrared and LiDAR data to detect the geothermal anomalies - an indicator of geothermal potential in the subsurface - in Indonesia. Assouline et al. [80] combined National Aeronautics and Space Administration (NASA) soil moisture and meteorological data to estimate three metrics related to geothermal energy potential: ground temperature, thermal conductivity, and thermal diffusivity. Using this approach the authors mapped shallow geothermal energy potential over the entire area of Switzerland. Lago and Rodriguez [78] estimated shallow geothermal energy potential using a combination of satellite and geological features to detect geographic regions with high levels of geothermal potential. In all of the aforementioned publications, the authors employed tree-based ML methods (e.g., Random Forests or Decision Trees) which are often favored due to their robustness to input data with diverse statistical properties (e.g., different numerical scaling and distribution).

*Hydroelectricity generation estimation.* A recent study [81] explored the use of RSD for estimating hydroelectric power generation (see Figure 5c). The authors use several explanatory RSD types (e.g., precipitation, soil moisture) and a Random Forest regression model to estimate the water discharge of the Shire river in Malawi, as well as concurrent changes in the electricity production at a local hydroelectric power station. The authors show that this regression model is highly predictive of daily discharge measurements when compared to discharge measurements taken from three ground facilities - termed gauge stations. The authors also report qualitative evidence suggesting that their model is predictive of daily hydroelectricity generation.

*Coal resource characterization.* REMOTE-ML techniques have been developed to estimate the quantity and type of coal, or the existence and size of existing coal mine (see Figure 5e). This information can be difficult to extract using traditional methods such as field measurements. Demirel et al. [82] monitored changes in the geographic size of coal mines over four years at a spatial resolution of 1 m over an area of 5000 $km^2$. Other researchers [83]–[87] used ML models to segment coal types (classified by different fixed carbon content in coal) [83], geology type (bituminous coal, lignite, sandstone, soil, coal gangue, marl, shale or other) [83], [85] . In addition,

Xiao et al. [87] performed coal detection in open pit coal mining areas in China and Vietnam using multi-spectral satellite imagery in approximately 220 km$^2$ area, at spatial resolution of 30m. Wang et al. [88] classified land use (e.g., as residential, coal mine, forest, grass, water, cultivated, unused and transportation) in coal mining areas in China at a resolution of 30m. Mukherjee et al. [89] detected opencast coal mine regions in India using color satellite imagery at a resolution of 30m.

*Oil infrastructure identification.* These REMOTE-ML approaches leverage computer vision algorithms to geolocate oil infrastructure captured in satellite and aerial imagery (see Figure 5f). Yao et al., Zhang et al., Wang et al., Zalpour et al. and Jivane et al. [90]–[94] used optical satellite images to detect oil tanks offshore across the globe. Song et al. [95] located oil wells based on satellite imagery in oil rich areas in China. Sheng *et al.* [96] used publicly available National Agriculture Imagery Program (NAIP) aerial imagery to identify the locations of oil refineries and petroleum terminals across the contiguous United States. Their findings include many instances of assets that were not captured in existing public databases, illustrating a key benefit of REMOTE-ML applications for addressing data gaps and enriching inventories of energy infrastructure.

*Accessibility and reproducibility.* Only 11% (2/18) of the publications in this literature published code for their methods. However, the data used for these studies were generally freely accessible due to the data generally being of lower resolution (coarser than 0.5 m/pixel).

### 4.2. Generation Infrastructure

#### 4.2.1. Electricity generation infrastructure identification and characterization

*Introduction.* In this section we discuss REMOTE-ML methods for identifying the location of and/or characterizing electricity generation infrastructure in a given region. Data about the location and properties of existing generation infrastructure are crucial for effective capacity planning, operations research, energy modeling, and policymaking. Conventional methods for mapping electricity generation infrastructure are often time-consuming, expensive, unreliable, or altogether infeasible. One common approach is to solicit this information from government agencies, utilities, or third-party operators; however, information may be spread over multiple organizations; or the data may not be comprehensive; and/or these organizations may be unable or unwilling to share the data. Another common approach to obtain infrastructure data is through field surveys, which can be completed by third parties (e.g., residents of the targeted geographic region), or by manual in-situ inspection of the targeted region. Surveys can be effective tools for collecting data that do not otherwise exist. However, they are time-consuming and expensive due to the manual effort involved in the process. Motivated by these challenges researchers have developed promising REMOTE-ML methods to automate collection of this information.

*Methodological Summary.* In these approaches ML models are typically trained to detect, segment, or classify the target energy infrastructure in the RS data. Energy information can then be inferred based upon the output of these ML models. For example, Figure 3(center) illustrates building and vehicle detection, where the model output consists of image coordinates for each detected object, from which the geo-locations can be inferred. Figure 3(right) illustrates the segmentation of buildings (among several objects) which returns the pixels corresponding to each building, from which their (approximate) area and energy consumption can be estimated. A variety of models have been used for REMOTE-ML applications for electricity generation infrastructure: Deep Neural Networks (DNNs) [4], Support Vector Machines [5], Random Forests [97], K-means [98], and regression [99]. In recent years, however, DNNs have led to substantial accuracy improvements on these tasks, and they are now the most widely used model [4], [100]–[103]. A wide variety of RSD has been employed with these models: color, infrared, multi-spectral, nighttime lights, and more. However, the majority of REMOTE-ML applications involve high-resolution optical overhead imagery.

*Energy Information: Rooftop Solar PV Arrays.* Mapping and characterizing distributed rooftop solar PV is becoming important as its deployment rapidly increases, however it remains technically challenging. REMOTE-ML methods have recently emerged to locate individual rooftop arrays [5], [97], [102], [104]–[106] estimate their size and shape [101], and estimate their energy generation capacity [99], [101], [107]. These methods have been employed in diverse geographic settings including the U.S, Europe [108], and Asia [109]. Moreover, these methods offer a tractable approach for automating the collection of relevant energy information over large geographic areas. Notably, the SolarMapper approach was used to identify individual rooftop arrays and estimate their generation capacities over the U.S. state of Connecticut [101]; and DeepSolar was used to count the number of rooftop arrays within small regions (e.g., census tracts) over the entire contiguous U.S. [107] and Germany [110].

*Energy information: Additional Infrastructure.* Methods for identifying a variety of other types of energy infrastructure have also recently been explored. Zhang and Deng [4] used publicly available Google Earth satellite data to find fossil fuel power plants in China. Zhou et al. [103] localized wind turbines (Figure 5m) from a satellite-based dataset provided by the U.S. Geological Survey (USGS) on wind turbines across the US. Abedini et al. [111] used UAV imagery to detect wind turbines in Iran.

*Accessibility and Reproducibility.* Only 17% (3/16) of papers on this topic published their code, making the reproducibility of these methods more challenging. On the other hand, many of the papers make use of the well-known deep neural network models, which are usually accessible via popular software packages such as PyTorch [112] and TensorFlow [113]. RSD sources needed to replicate these methods are usually accessible, with 83% (14/16) of publications using publicly available data source (public satellite imagery) and/or Google Earth/Engine/Maps, the other 17% not providing clear details about their source of RSD [109], [111], [114].

### 4.2.2. Electricity generation infrastructure risk monitoring

*Introduction.* In this section we discuss approaches for monitoring potential risks to energy generation infrastructure posed by physical deficiencies; such as detecting solar array hot spots or dust accumulation, wind turbine blade defects, and hydroelectric dam structural defects (see Figure 5(j-o) for illustrations). Effectively monitoring these facilities can reveal infrastructure failures in real-time so as to minimize costs, and/or support effective maintenance to improve the lifespan and efficiency of the infrastructure. Conventional methods for monitoring usually involve manual visual inspection of the infrastructure, which often requires extensive time and human effort, and may be error-prone. For example, solar and wind farms often include large numbers of individual panels and turbines, respectively, that must each be inspected on a regular basis. Furthermore, these infrastructure assets are often distributed over large geographic regions that may also be remote. In some monitoring tasks, the energy infrastructure may also be difficult or dangerous to access; e.g. inspecting the walls of dams for cracks. Motivated by these challenges, researchers have investigated the use of REMOTE-ML approaches to automate the monitoring of energy infrastructure for risk purposes (Figure 5(m-o)). In principle, these approaches can dramatically reduce the amount of time and human effort required for monitoring. Additionally, whereas manual inspection is onerous and error-prone, automated methods can monitor continuously without exhaustion-related or attention-related errors.

*Methodology Summary*: Nearly all of the REMOTE-ML approaches in this section rely upon the remote collection of color or infrared imagery over the target energy infrastructure. Then, ML models are trained to recognize visual cues on the infrastructure that indicate a condition of interest (e.g., a malfunction, or damage). Some examples of this strategy include the identification of damaged lightning receptors on wind turbines [115], dust on a solar panel [116], or cracks on dam surfaces [117], [118]. Due to the small scale of the visual features that are usually of interest, nearly all methods rely on UAVs or aerial photography [117], [119] as their input, due to the ability of these platforms to provide especially high-resolution imagery (e.g. much higher than < 0.3 m achievable with satellites) with various bands (e.g., color [120] or infrared [121]. Since most of the remote sensing data in these tasks takes the form of imagery, REMOTE-ML methods usually rely on DNNs due to their state-of-the-art image recognition accuracy [115], [118], [122], [123]. One paper explored the use of Support Vector Machines [116].

A large portion of the work on this topic is focused on the inspection of solar panels, and therefore we divide our energy information summary into two sections: solar panel monitoring, and all other topics, which primarily include wind turbine and hydro station monitoring.

*Energy Information: Solar Panel Monitoring.* Multiple studies [119], [121], [124] identified damaged solar panels in Europe and China, producing bounding boxes on damaged panels or smaller damaged hotspots in the imagery on large solar arrays (typically seen in solar farms). Some studies [116], [120], [122], [125] detected and classified solar panel anomalies into several different categories related to the types of potential hazards: dust-shading, corrosion, yellowing, snail trails, encapsulant delamination, and glass breakage. If none of these were detected, the panel was otherwise estimated to be in good condition. Hanafy [126] estimated the level of solar panel cleanliness (portion of PV aera being covered by dust) in a laboratory setting for a single solar panel.

*Energy Information: Wind Turbine and Hydro Station Monitoring.* Wang and Zhang [127] detected wind turbine cracks in China. Moreno [128] did not use real-world imagery, but performed a proof-of-concept study using simulated remotely-sensed data, showing the potential capability to detect blade anomalies in real-world remotely-sensed data. Some studies [115], [123], [129] operated UAVs anomalies on wind turbine blades including scratches, shedding, sand holes, cracks, and seepage erosion. Wang [130] proposed an unsupervised learning scheme to address the scarcity in labelled training data of wind turbine blade anomaly in Japan. Regarding hydro power station inspection, Li [118] and Feng [117] both developed methods to detect surface cracks on dams.

*Accessibility and Reproducibility.* Only 10% of the studies described here published software to replicate their methods. This is not a major obstacle however, since their methodology was described in sufficient detail for replication, and most of the methods are easily implemented using public (e.g., Python's sklearn [131] or scikit-image [132]) or proprietary (e.g., MATLAB[2]) software packages. By contrast, the explanatory RSD represent a significant obstacle to replication of existing work, and to a lesser extent, application of the methods in new contexts. Only three of eighteen publications utilized publicly-available RS data, or otherwise publicly released their datasets. Most of these studies rely upon UAV imagery or aerial photography, whose accessibility depends and varies strongly across political boundaries.

### 4.3. Transmission Infrastructure

#### 4.3.1. Transmission infrastructure identification and characterization

*Introduction.* Electricity travels from generators along larger, high-voltage transmission and smaller, low-voltage distribution lines to reach end users. The higher voltages reduce line losses and improve the efficiency of the system as electricity is transferred over large distances. Generators connect to high voltage transmission lines via transformers (which increase the voltage to higher levels) at electric substations. These voltages are then reduced at substations near the destination, to enable connections to the lower voltage distribution system. Distribution lines, consisting of medium and low voltage networks, then deliver electricity at a reduced voltage level to end users for residential or business usage. Transmission and distribution lines form the power grid and provide electricity across large geographic regions. While high-voltage line data are often available, low- and medium-voltage transmission line data, critical for responding to electricity access and climate change challenges, are often unavailable due to the lack of a systematic collection of grid information [133]. Due to these difficulties, researchers have recently explored the possibility of using machine learning models to automatically locate transmission and distribution lines in RS imagery. These methods often identify related objects as well, such as the towers and poles that support the power lines, or vegetation that may be encroaching on the lines. Using this information it is then possible to produce a detailed map of the power grid over large geographic areas.

*Methodological Summary.* These REMOTE-ML methods usually rely upon visually identifying transmission infrastructure (Figure 5p) in color or infrared imagery, or LiDAR point cloud data, collected from airborne platforms (mostly UAVs, some helicopters and surveillance cameras). Airborne platforms can capture much higher resolution imagery than other remote sensing technology. Moreover, vertical overhead imagery of transmission infrastructure can be occluded (e.g.,by tree canopy), limiting the effectiveness of satellites, while airborne platforms can navigate to capture more favorable imagery. ML models are used to automate the localization of the transmission infrastructure in these imagery. Because of their excellent performance on visual recognition tasks, DNNs have underpinned most recently-proposed methods to perform this task. Several other ML models have also been explored as well; Support Vector Machines, Decision-Tree-based methods, and others.

---

[2] MATLAB is a software product by MathWorks®

*Energy Information:* A major difference among proposed methods is the spatial precision with which they localize the transmission infrastructure. For instance, some methods annotate objects present in an image with rectangular boxes, while others label each *individual pixel* in an image as either residing on an object of interest, or not. Examples of specific implementations of bounding box annotation include: power line cable detection [134]–[137], invading foreign object detection [138]–[140], and transmission tower, distribution pole and other component detection [141]–[147]. As for pixel-level labeling, it has been applied to power line cables [148]–[160]. Among the body of work we reviewed, most studies focused on applying similar methods on the aforementioned objects of interest, rather than technical improvements of the methods. While many papers don't specify the location of their work, a variety of geographic regions are covered by the searched literature. Experiments were carried out in Asia [134], [138], [139], [141], [143], [147], [150], [160]–[163], covered some regions in Europe [144], [146], [151], [154], [159], [164] ; worked in Oceania [165], [166], in North America [155], [156], and in South America [167].

*Accessibility and Reproducibility.* Assessing the generalizability and comparability of methods in this space is challenging because these applications rely on airborne imagery that is often procured for private use in a specific study. Although some datasets used in the studies cited were made public [134], [152, p. 20], [156], the inability to reproduce results remains a major challenge in this space as most papers still used different private datasets and no datasets were identified as a benchmark to test method performance.

### 4.3.2. Transmission line health monitoring

*Introduction.* Transmission lines and their associated infrastructure are susceptible to equipment malfunctions and impacts from the encroachment of vegetation within the transmission corridor. The failure in any part of the transmission and distribution grid may cause imbalances and therefore instability in electricity supply for a region [168]. To mitigate the risks associated with the hazards, routine inspection is required to monitor the growth of vegetation and other hazards along the transmission corridor. However, transmission lines can stretch across long distances, traversing remote and hard-to-reach places, making manual inspection labor-intensive and extremely difficult [168]. Emerging remote sensing technologies, particularly UAVs combined with ML analysis enable fast, automated monitoring of transmission corridors. Below, we detail ML applications in this space (Figure 5q) that have been shown effective in executing inspection tasks that were conventionally performed manually.

*Methodological Summary.* For the purpose of transmission line health monitoring, both the infrastructure itself and the corridor environment have been focused on and two major groups of studies formed naturally based on the objects of the studies and the types of energy information extracted: (1) transmission line infrastructure characterization and (2) corridor environment mapping. The goal of transmission line infrastructure characterization is to assign a semantic label ("normal" or "breakage") to a given image, or the individual objects in that image (normally transmission towers or cables). Corridor environment mapping focuses more on objects in the nearby environment instead of transmission infrastructure itself and assigns a label to each pixel in an image from a group of objects present in the corridor environment (e.g. tree types, towers/pylons, buildings, grass, soil, river, sand, wires) along the transmission corridor.

Most studies of transmission corridors and health monitoring are based on UAV or aerial imagery and LiDAR is also frequently used for corridor environment mapping studies. Almost all datasets used in the studies we reviewed are private and no benchmark dataset was identified. ML models were trained to perform both tasks and the common methods used include convolutional neural networks, neural networks [145], [169], support vector machine [6], [140], [152], [170], [171] and random forests [172]–[174].

*Energy information: transmission Line Infrastructure Characterization.* Object characterization tasks for transmission line applications can be grouped into two categories. Some authors proposed methods that characterize entire images which contain multiple transmission line cables. For example, Zhang et al. [140] classify images by the presence or absence of foreign objects anywhere in the entire image; Wang et al. [170]) classify entire images into "good condition" or "fault condition". Other papers classify transmission line objects which usually are output of a preceding localization step and return one label for each object. (Mao et al. [152] classify various fault types of transmission line cables ("normal", "strand breakage", and "foreign matter"). Lin et al. [175] classify the icing condition types of transmission lines ("no icing", "slightly icing", "moderate icing", "severe icing", "dangerous icing" and "fault warning icing"). Sampedro et al. [145] classify four types of transmission towers that are different in shapes and sizes. In terms of the geographic coverage of the work, Lin et al. [175] carried out experiments in Asia while other papers didn't specify their geographic coverage.

*Energy information: Corridor Environment Mapping.* In 2010, Li et al. [171] fused LiDAR and color images to classify among three different tree species and power lines along the transmission lines with resolution of 15cm. Frank et al. [6] used a similar approach combining LiDAR and hyperspectral images from UAV to segment the imagery into one of twelve categories (ranging from transmission towers, to a variety of tree types) with a spatial resolution of 50cm. Others [169], [172]–[174] solely used LiDAR data to segment the transmission corridor scene into categories such as transmission lines, vegetation, building and ground with 15cm resolution. Subsequently in 2018 Guo et al. [176] segmented transmission line surroundings into six classes (tower, water, woodland, grassland, bare land, buildings) at 5cm resolution. Qayyum et al. [177] explored a unique approach with a stereo camera (binocular) mounted on a UAV and used a DNN to predict the object (mainly trees) height directly with 15cm resolution.

*Accessibility and Reproducibility.* None of the papers provided a public code base to replicate their work, and very few of these papers used publicly available data or made their datasets publicly available.

### 4.3.3. High-voltage Insulator Monitoring

*Introduction.* High-voltage insulators are electrical hardware that are typically installed in high-voltage electricity transmission systems to provide mechanical support for, or separation among, conducting electrical components (e.g. power lines) without themselves conducting electricity. Insulators may be prone to sudden failure, leading to power outages as well as damage of expensive electrical infrastructure [178]. A common practice to mitigate risks of failures is to regularly conduct manual visual inspections of the insulators. However, this process is time-consuming and expensive because there are typically large numbers of insulators installed over large, and sometimes remote, geographic regions. Furthermore, manual human inspection can be error-prone due to the laborious and repetitive nature of the inspection task [179]. This has motivated the development of REMOTE-ML methods that can automatically inspect the insulators using video or imagery from UAVs. This approach (Figure 5r) is potentially much faster, cheaper, and more reliable than manual inspection.

*Methodological Summary.* Automatic insulator inspection using ML is usually broken down into two major steps: (i) localization and (ii) characterization. The goal of localization is to determine the precise location of each insulator in a given UAV image. To accomplish this, a machine learning model can be trained to automatically assign a rectangular box (or other type of annotation) to each insulator in a given input UAV image. The goal of characterization is to assign a semantic label (e.g., "healthy" or "risky") to a given UAV image, or the individual insulators in that image. Sometimes a single label is assigned to the entire image which may contain multiple insulators (e.g. Zhao et al. [180]); however, it is more common to use the output of the localization step to separately analyze and assign a label to each insulator present in the image – a richer output (e.g. Zhao et al. [181]). Similar to localization, a machine learning model can be trained to assign semantic labels to insulators automatically. Collectively, these two steps can be used to produce maps of insulators and their respective levels of risk. In turn, this information can be used to prioritize subsequent manual inspections and insulator replacements to maximize efficiency in identifying potential at-risk insulators.

Historically, a wide variety of ML methods have been explored to perform insulator inspection: Support Vector Machines [179], [181], [182]. Boosted Decision Trees [142], [183], [184], and others [183]–[187]. In recent years, DNNs have also led to increases in the accuracy of insulator mapping, and now underpin most proposed methods [178], [181], [188]. As input, automatic insulator inspection relies upon color or infrared imagery collected from UAVs (e.g., drones) or airplanes. The primary motive for using UAVs is their ability to capture higher resolution imagery than other remote sensing technology, which is needed to recognize the fine-grained visual details of individual insulators. We next review the types of energy information that have been extracted using these techniques. We divide our review into two groups: (1) localization information and (2) characterization information while also noting that studies frequently performed both of these steps, and therefore appear in both groups.

*Energy Information: Insulator Localization.* The spatial precision of localization varies from localizing individual insulator units to localizing strings of insulators across the methods proposed in the literature. The authors [187], [189], [190] proposed methods to automatically annotate individual insulators present in an image with rectangular boxes. In high-voltage transmission lines, multiple insulators usually appear together in tightly-spaced groups called strings, and the methods [162], [178], [182], [188]–[205] were all designed to annotate strings of insulators with boxes, rather than individual insulators. Another body of work has focused on developing methods that label each pixel in the input image as either residing on an insulator, or not [179], [186], [187], [192], [194], [196], [206]. This pixel-wise labeling, also known as segmentation, has the advantage that it provides more precise size and shape information about the insulators, but is also considered a more difficult task as the output needs to be more accurate and the creation of human-labelled ground truth is more difficult.

*Energy Information: Insulator Characterization.* Most characterization methods take the output from a preceding localization step to isolate and separately analyze the imagery corresponding to individual insulators or a string of insulators. As a result, these methods return one class label for each insulator [181], [183]–[185], [187], [189], [190, p.], [194], [207], or string of insulators [187, p. 2019], [189], [197]. In contrast to most studies, Zhao et al. [180] provide image-level characterization where a condition is assigned to an entire image that contains a string of multiple insulator units. Similar to the literature on localization methods, most work on characterization has focused on improving the algorithms for automatically characterizing insulators in an effort to increase the accuracy of output. However, one major source of variability among recent work is the semantic labels that the ML models assign to the insulators. Semantic labels are defined differently in different studies and we enumerate most of them here: Zhao et al. [181] classify insulators into several conditions including "normal", "crack", "defect", and "dust"; Prasad and Rao [185] classify insulator units as "healthy", "marginal" or "risky" conditions; Potnuru and Bhima [207] characterize insulator units into "healthy" or "risky" conditions; Liu et al. [189] characterize insulator strings into voltage levels (220 kV, 330 kV, and 550 kV); Ling et al. [194] characterize malfunctioning insulator units ("self-blast", as defined by the authors); Pernebayeva et al. [183], [184] characterize insulator units by their icing level ("clean", "water", "snow", or "ice"); Jalil et al. [197] classify insulator strings as "normal" or "rust" conditions; Sampedro et al. [187] classify insulators as "damaged" or "missing"; and Antwi-Bekoe et al. [186, p.] characterize insulators as "defective" or "not defective".

*Accessibility and Reproducibility.* UAVs are also now commercially available at relatively low costs, making it possible to design custom flight paths in permitted areas and collect high-resolution imagery on demand. This makes automatic insulator inspection techniques accessible to researchers and practitioners. Unlike public satellite or aerial imagery that is standardized, UAV imagery is often collected on demand for private use. Furthermore, the lack of documentation on the spatial resolution and geographic coverage in private data collection makes it difficult to compare results across studies, as there are no publicly-available benchmark datasets on which all methods have been compared.

### 4.4. Electricity Consumption

#### 4.4.1. Building energy consumption

*Introduction.* Information on energy consumption over a geographic region is crucial for regional infrastructure planning and maintaining balance between energy supply and demand [208]. Building energy consumption accounts for a significant share of total electricity usage and greenhouse gas emissions [209]—residential and commercial buildings, alone, accounted for roughly 40% of U.S. emissions in 2019 [210]—and is the most represented sector of energy consumption in the body of work that we reviewed. However, characteristics and individual measurements of building energy consumption available to utilities are typically unavailable to the public and researchers due to their security and privacy considerations [208]. REMOTE-ML methods provide an alternative (Figure 5s) to acquire such information and have been utilized to estimate energy consumption over city, province, or even country level.

*Methodological Summary.* The most common modality of RSD used for estimating energy consumption was nighttime lights (NTL) imagery [8], [208], [211]–[216]. NTL imagery records the radiance (e.g., watts per unit area) from the earth in the near-infrared and visible bands during nighttime hours. Many types of human activities involving electricity produce NTL (e.g., street lighting, households) [217]–[219], making it possible to infer both electricity access and reliability. For example, NTL pixel intensities have been shown to be correlated with electricity access, while variability in the NTL intensity over time has been shown to be correlated with lower electricity reliability. NTL intensities have also been shown to be predictive of building electricity consumption, which is a significant part of electricity consumed for lighting [8]. We note that NTL have been used to infer a variety of other useful information (see Levin et al. [220]; Zhao et al. [221]) but we focus here on energy consumption estimation. NTL data used in these papers mainly come from two sources: Visible Infrared Imaging Radiometer Suite's Day/Night Band (VIIRS-DNB) and U.S. Air Force Defense Meteorological Satellite Program Operational Linescan System (DMSO-OLS).

In addition to NTL, other RSD modalities have been used to estimate energy consumption. Prominent examples include building footprints data [222], Normalized Difference Vegetation Index (NDVI) [209], [223], [224], Land Surface Temperature (LST) , and high-resolution visible spectrum imagery [222]. These RSD have also been used together with non-RSD, such as census data [209] and features of the built environment (e.g., the size and age of buildings; or building roof shapes [224], [225]). The majority of methods for estimating energy consumption use some form of regression. Both linear and non-linear models (e.g., Neural Networks) have been used for the regression task. However, in some of the work we reviewed, classification models (including DNNs [222] and Random Forests [224]) were also used to extract feature variables to be used in energy consumption regressions.

*Energy Information: Energy Consumption Estimation Based on Nighttime Lights (NTL).* A number of studies in this space estimate energy consumption utilizing NTL but tend to vary in the geographic coverage and resolution of their response (output) data. Xie et al. [211] reported sub-city level estimation in China where urban and suburban areas were studied separately. Sahoo et al. [216] reported sub-state level estimation in India where the State of Uttar Pradesh was divided into 19 zones based on their respective power suppliers. Townsend et al. [208] was the earliest work to our knowledge to study the relationship between NTL and state-level energy consumption, and explored this relationship in Australia. At a similar spatial resolution, Fehrer and Krarti [213], and Jasinski [8] estimated energy consumption using NTL at the city, state, and county level, respectively. Some other studies performed estimation on a country level, estimating energy consumption in Tanzania [212] and India [214]. Lu et al. [215] did it at a lower geographic resolution where estimation was conducted on country, continental, and global scales.

*Energy Information: Energy Consumption Estimation Using Other Remote Sensing Data.* A large body of work uses RSD sources other than NTL to estimate energy consumption. Some of these papers estimated energy consumption at a spatial resolution as high as each building. Streltsov et al. [222] extracted size and shape of buildings from 1m resolution aerial imagery and estimated energy consumption at the resolution of individual buildings in two cities in the U.S. Faroughi et al. [225] estimated energy consumption at building level using LST coupled with other physical variables of the built environment obtained from literature (e.g. housing type, housing size) in Iran. Chen et al. [223] estimated building level energy consumption using LST and NDVI and a variety of other indicators describing urban morphology (e.g. urban layouts, urban geometry, dwelling types) in an urban area in Eindhoven, the Netherlands. Other studies estimated energy consumption at a lower resolution: the block/neighborhood level. Ma et al. [209] used NDVI and other socio-economic metrics to estimate building energy use in New York City, USA at block level. Garbasevschi [224] used NDVI coupled with other building attributes collected from non-remote-sensing data sources to first classify building ages and then further estimate building energy demand in Germany at neighborhood level.

*Accessibility and Reproducibility.* All papers we reviewed in this section have thorough documentation of what data sources they used and most of the data sources are publicly available. However, none of the papers provided the code base or other implementation details for reproduction. Therefore these methods are accessible, but may be difficult to reproduce exactly.

### 4.4.2. Electricity Access and Reliability Estimation

*Introduction.* One in ten people around the globe lacked access to electricity in 2019 [226], and access to modern electricity service is correlated with quality of life [227]. Electricity access information is useful for assessing electrification progress, or planning for future investments in extending electricity access [219], [228], [229] through either grid extension, microgrid development, or off grid distributed generation like solar [230]. In this section, we discuss methods to estimate whether people living in a certain geographic location (e.g., a town, or village) have access to reliable electricity. While there are different levels of electricity access [230], electricity access estimates have traditionally been reported as a binary number, indicating whether a given region has access, or as an electrification rate indicating the proportion of individuals that have access. Researchers have also recently become increasingly interested in measuring the reliability of the electricity supply [231], motivated by the hypothesis that low energy reliability may undermine the benefits of electricity access [81]. This would make reliability an important complementary measure in many decision-making processes involving access. Measures of reliability vary, but they generally attempt to encode the consistency of the electricity supply in a region. For example, Bhatia [230] reported the average hours each day when electricity was unavailable. Historically, and predominantly today, electricity access and reliability estimates have been obtained using household surveys, which are labor and time intensive, and must be repeated regularly for updated information [228], [232]. This has motivated the development of REMOTE-ML methodologies (Figure 5t) to automate these assessments.

*Methodological Summary.* Although electricity access cannot be measured directly from RSD, it can be inferred by observing human activity in RSD that is correlated with electricity access. One such example is nighttime lights (NTL), which forms the basis of most energy access estimation approaches. The mechanism by which NTL can be used to infer energy access, reliability, and other human activity is discussed in more detail in the Methodological Summary of Section 4.4.1 above. For energy access and reliability estimation, ML models are often trained upon pre-processed NTL imagery, or more complex statistics computed on the NTL imagery. Although many models operate directly on NTL imagery it is important to note that the raw NTL imagery undergoes a substantial level of pre-processing (e.g., cloud removal, background noise removal) before the ML models are applied. The methods explored thus far utilize two different NTL sensors: the DMSP-OLS satellites [233], and the more recent VIIRS satellites [234].

*Energy Information: Electricity Access.* Early work on this topic by Doll et al. [235] used population maps and NTL imagery to estimate the electrification rates of rural populations within *every* nation-state globally. Subsequent work in Min et al. [217], [219] used NTL imagery to make binary estimates of whether individual villages in Mali, Senegal and Vietnam were electrified. Dugoua et al. [236] used NTL imagery to estimate the number of electrified households in individual villages, producing estimates for 600,000 villages spanning all states in India. Most recently Falchetta et al. [228] created a publicly-available dataset of energy access estimates (and energy consumption levels), which is notable because it is built upon regularly-updated and publicly available population and NTL data. As a result, the authors suggest that the dataset will be updated regularly as new explanatory data becomes available. Falchetta et al. [232] utilize the aforementioned dataset to estimate national and sub-national electrification rates in sub-saharan Africa. They also evaluate the changes in electrification rates over time across Sub-saharan Africa to reveal inequalities in the pace of electrification. Räsänen et al. [237] used a similar overall methodology to estimate the per capita electricity consumption at a village-tract level in Myanmar.

*Energy Information: Electricity Reliability.* Alam [238, p. 20] first showed NTL imagery might be predictive of electricity reliability by demonstrating that NTL statistics were correlated with (i) power outages and (ii) the percentage of electricity shortfall at peak demand, at the state level over India. Min et al. [239] found that NTL statistics were correlated with both the number and duration of electricity interruptions, in rural regions of India. Two studies also explored the ability to estimate the reliability of electricity access using NTL imagery, rather than simply establishing correlations. Mann et al. [240] trained a Random Forest model to make daily predictions of whether a power outage had occurred at 65 sites in Maharashtra, India. Cole et al. [241, p. 2] used a Neural Network to make daily predictions of the number of power outages within 1 km² regions in the Northeastern US, spanning a 25,000 km² area.

*Accessibility and Replicability.* Nightly global imagery from VIIRS (the explanatory RSD for most recent studies) is publicly available, making energy data extraction using NTL imagery accessible to researchers. In particular, much of the explanatory data needed to apply these methodologies in new contexts is public, and available globally. In particular the dataset published by Falchetta et al. [228] provides processed NTL imagery and population maps for the entire globe, and is projected to be updated annually, facilitating exploration of these methods. There are also a number of more advanced NTL satellite systems being deployed, with data of varying levels of accessibility - see Levin et al. [220] for details. However, replication or adoption of existing methods is made somewhat difficult by the large quantities of data that are often involved (e.g., see Min et al. [239]), as well as the absence of published software implementations of the methods.

## 5. SYNTHESIS: CHALLENGES, OPPORTUNITIES, AND RECOMMENDATIONS

After detailing the state of REMOTE-ML research in each application area described in the previous section, in this section, we critically analyze REMOTE-ML as an emerging field of research and identify crosscutting trends observed across application areas within the REMOTE-ML literature. We identify six major emergent themes. For each theme, we proceed with an overview of the topic area and also highlight key trends and limitations. Lastly, we discuss opportunities for improvements in REMOTE-ML techniques, future research and application areas, and offer recommendations to accelerate innovation.

### 5.1. Model trustworthiness: robust characterization of REMOTE-ML accuracy

REMOTE-ML methods are commonly used to generate energy data for use by energy researchers and decision-makers, and therefore the error of REMOTE-ML-derived output data is a key consideration. Inaccurate assessment of this error can propagate through subsequent modeling and analysis workstreams, potentially limiting the credibility of insights derived from them. Additionally, without reliable error measures, objective comparison of REMOTE-ML techniques is impractical. Unreliable error estimates of REMOTE-ML output data can be caused by improper evaluation of the underlying ML models. Below we recommend a set of best practices to support robust evaluation of REMOTE-ML methods.

After training a ML model its ability to produce reliable insights must be evaluated empirically on a collection of testing data. Reliable error estimates can generally be achieved by adhering to the following principle: *use a testing dataset that faithfully reflects the data that will be encountered in the model's intended end-use application*. Failure to use representative, application-relevant test data for model evaluation may lead to unreliable—and often overly optimistic—error estimates. This pitfall can be avoided by applying the following two corollaries: (1) use independent testing data and (2) use data that are representative of the model's desired application.

*Use independent testing data*. The intended end-use of most REMOTE-ML methods is to apply it to RSD that was not used to train it. To reflect this end-use condition, the testing dataset for any particular ML model should be completely disjoint from the data used to train it. This can be achieved by using well-established techniques in the ML literature, such as splitting the total available data into two or more disjoint partitions (e.g., training and testing partitions), or by using cross-validation [242], [243]. Failure to do this can lead to positive bias (i.e., optimism) in the estimated error metrics [242], [243].

*Use testing data that are representative of the end-use application*. To ensure unbiased REMOTE-ML method evaluation, another key principle is to ensure that the testing data are sufficiently representative of the method's intended application. For example, if the intended use is to detect solar PV arrays in California, then the testing data should be representative of the variety of conditions and physical settings likely to be captured by RSD data across the state of California. Similarly, if the goal is to develop a method that can robustly detect solar PV arrays across the globe, then the testing data should be representative of the variation in RSD that would be encountered globally; a much more demanding (and often unattainable) requirement. Besides geographic location, there are many factors that feed into representativeness such as scene content (e.g., types of vegetation, or features of the built environment), time of day, weather, and imaging hardware. ML models can be sensitive to these variations in the RSD, with substantial impact on their error rates [244]–[246]. Error rates gauged on testing data may only be valid for model applications that use RSD possessing similar characteristics as the test set. In practice, it is often difficult to collect sufficiently diverse explanatory RSD that encompasses the full spectrum of variation that might be encountered at end-use. In these cases, the reported error metrics are most reliable with respect to RSD that were collected under similar conditions to the test data, and may vary substantially on data collected under other conditions.

While the discussion here focuses on robust *evaluation* of accuracy, we also note that ML models perform best when they are trained on data that is representative of the conditions that will be evaluated during testing, and encountered during subsequent application [242], [243]. Therefore, it is recommended that *training datasets* be representative of the conditions likely to be experienced during model application as well; this practice will ensure that model development and evaluation procedures are sufficiently aligned with their end-use needs.

*Trends and limitations:* Table 4 divides the REMOTE-ML literature into topical sub-areas, and the type of validation procedure used to evaluate their accuracy. The large majority of REMOTE-ML publications (77% overall) evaluate their accuracy, and do so on an independent testing dataset (i.e., k-fold cross validation or a train/test split), following best practices of the ML community [242], [243]. However, most publications report performance on a limited set of RSD conditions (e.g., geography, time-of-day, weather). Thus, although the model may perform well on these specific conditions, it is unclear how its performance would vary when applied to novel settings. Moreover, some publications do not provide precise details about the conditions under which their RSD was collected, and few publications clearly articulate the expected limitations of model performance.

*Opportunities and recommendations:* Obtaining reliable accuracy estimates for REMOTE-ML-derived energy data will be crucial to building trust in the methods, and translating them into viable tools for downstream modeling and decision-making. We recommend that researchers precisely describe the construction and composition of their testing datasets, and clearly articulate the conditions under which their error metrics are likely to be reliable indicators of performance. This will be increasingly crucial to providing realistic performance expectations and cultivating trust in REMOTE-ML techniques, especially as the community becomes more open; for example, when creating and taking advantage of shared repositories and collections of pre-trained models, as is common in other research domains. Because models that were developed under various circumstances and for different geographies will have varying relevance to specific energy applications, understanding the generalizability of each technique will be key to accelerating REMOTE-ML progress.

| Category | Total | Validation method | | | | |
|---|---|---|---|---|---|---|
| | | K-fold Cross Validation | Train/Test Split | Fit to Training Data | Qualitative | None |
| Solar Potential Energy Estimation and Forecasting | 34 | 7 (20%) | 23 (68%) | 0 (0%) | 4 (12%) | 0 (0%) |
| Wind Resource Estimation and Forecasting | 9 | 5 (55%) | 4 (45%) | 0 (0%) | 0 (0%) | 0 (0%) |
| Other Primary Resources: coal, geothermal, hydroelectric, and oil | 18 | 3 (17%) | 11 (59%) | 1 (6%) | 1 (6%) | 2 (12%) |
| Electricity Generation Infrastructure identification and characterization | 16 | 3 (19%) | 13 (81%) | 0 (0%) | 0 (0%) | 0 (0%) |

| | | | | | |
|---|---|---|---|---|---|
| Electricity Generation Infrastructure Risk Monitoring | 16 | 2 (12%) | 12 (75%) | 0 (0%) | 1 (6%) | 1 (6%) |
| Transmission and Distribution Line Mapping | 31 | 9 (30%) | 14 (45%) | 0 (0%) | 8 (25%) | 0 (0%) |
| Transmission infrastructure mapping & Monitoring | 12 | 4 (33%) | 5 (41%) | 0 (0%) | 0 (0%) | 3 (25%) |
| High-voltage Insulator Monitoring | 29 | 3 (10%) | 21 (72%) | 0 (0%) | 4 (14%) | 1 (3%) |
| Building Energy Consumption | 13 | 4 (30%) | 0 (0%) | 8 (62%) | 1 (8%) | 0 (0%) |
| Electricity Access and Reliability Estimation | 11 | 2 (18%) | 0 (0%) | 5 (45%) | 0 (0%) | 4 (36%) |

*Table 4: Breakdown of publications by the validation method used to measure the accuracy of their predictions.*

## 5.2. Replication: access to data (RS imagery and labels) and code

Replication challenges are not limited to work in REMOTE-ML, but are part of the larger conversation about ML methodological reproducibility [247]. In the ML context, reproducibility can be viewed as the ability to produce the same or statistically similar results from a method for a given set of data, or to be able to come to the same inferential conclusions based on a new instantiation of the experimental design [248]. Data sharing and code sharing, while not historically a critical part of research, has increasingly become desirable, if not required, in some fields to facilitate enhanced reproducibility. This has been facilitated through publisher requirements [249], and norms in the field [250]. While replication is a uniformly positive characteristic of research for ensuring reproducibility, there are also other ethical questions around data security and privacy that arise that will be discussed in a later section.

*Trends and limitations:* Roughly half of the reviewed studies (44%) did not publish or share their explanatory and response data. In Table 5, we report the percentage of studies in each topical sub-area that make their explanatory and response datasets accessible. In some cases, the explanatory data may be available (e.g. publicly available imagery), but the ground truth labels for the data are not shared, substantially increasing the efforts that would be required for researchers to independently verify or replicate the study.

The sharing of code bases that deal with the processing of RSD and development of machine learning models can accelerate progress in this research domain. As shown in Table 5 however, we found that just 6% of studies shared their code, and no studies in several topical sub-areas shared their code (e.g., Solar Potential Energy Estimation and Forecasting, Building Energy Consumption). Not sharing the code for a study makes replication and adaptation difficult. One noteworthy exception to these trends was Falchetta et al. [228] who published their explanatory and output data on sub-Saharan Africa electrification as a research deliverable, along with software to easily reproduce their experiments and results.

One additional limitation of a lack of data sharing is the lack of benchmark datasets. Without benchmark datasets, direct comparison of algorithm performance that fairly evaluates their relative strengths is not feasible since the algorithms would be tested on different datasets. For this reason, we have refrained from making direct comparisons of accuracy of the REMOTE-ML models in this study because too few had experimental designs similar enough to enable fair direct comparisons.

*Opportunities and recommendations:* Establishing data and code sharing for every study as a norm will enable significant increases in reproducibility, and will facilitate direct comparison of these approaches to other methods, encouraging constructive competition and research innovation. Journal requirements around code sharing and data availability would promote adoption of these practices among the research community.

| Category | Total number | % provide code | % publish or use publicly-available, in both explanatory and response data | | |
|---|---|---|---|---|---|
| | | | Yes | Maybe | No |
| Solar Potential Energy Estimation and Forecasting | 34 | 0 (0%) | 16 (47%) | 8 (23%) | 10 (30%) |
| Wind Resource Estimation and Forecasting | 9 | 0 (0%) | 6 (66%) | 2 (22%) | 1 (11%) |
| Other Primary Resources: coal, geothermal, hydroelectric, and oil | 18 | 2 (11%) | 2 (11%) | 0 (0%) | 16 (89%) |
| Electricity Generation Infrastructure identification and characterization | 16 | 3 (19%) | 10 (63%) | 0 (0%) | 6 (37%) |
| Electricity Generation Infrastructure Risk Monitoring | 16 | 2 (13%) | 1 (6%) | 0 (0%) | 15 (94%) |
| Transmission and Distribution Line Mapping | 31 | 1 (3%) | 5 (16%) | 0 (0%) | 26 (84%) |
| Transmission infrastructure mapping & Monitoring | 12 | 0 (0%) | 0 (0%) | 0 (0%) | 0 (0%) |
| High-voltage Insulator Monitoring | 29 | 0 (0%) | 0 (0%) | 0 (0%) | 0 (0%) |
| Building Energy Consumption | 13 | 0 (0%) | 7 (53%) | 0 (0%) | 6 (47%) |
| Electricity Access and Reliability Estimation | 11 | 3 (27%) | 1 (9%) | 6 (54%) | 4 (36%) |

*Table 5: Percentage of publications that make their code publicly available ("% provide code") and the percentage of publications that make their explanatory (input) and response (output) data available.*

### 5.3. Application gaps and opportunities in existing work

This systematic review discusses numerous applications related to energy systems and what can be identified and evaluated through remotely sensed data. Here we identify gaps (and corresponding opportunities) in the existing literature with respect to the types of energy data that have been extracted, and their geographic coverage.

*Trends and limitations:* Our review revealed that the majority of work in the REMOTE-ML space has been focused on three primary areas: (1) energy resources, (2) power system infrastructure and assets, and (3) energy access and end-use consumption. Among these areas, we identified very few applications outside of the electric power sector that have received significant attention. For instance, we identified few (7 out of 185) publications dealing with oil and gas exploration, production, or transportation; even though oil and gas resources, and their applications in transportation, account for 57.4% of global energy consumption [251]. Note that our review excluded papers on environmental impacts, such as methane leakage or oil spill detection, which were outside the scope of this review.

A compelling promise of RSD is the unique possibility it offers to produce actionable insights at scale through earth observation. Interestingly, although many sensors have broad geographic coverage, we found that the vast majority of the work reviewed here was conducted for fairly limited geographic sizes, with a few notable global or national exceptions such as Arderne et al. [133] or Falchetta et al. [228]. In practice, although relevant RSD may be available for global applications, scaling up REMOTE-ML techniques requires additional research to develop and implement robust models whose performance is generalizable across diverse regions.

Furthermore, we identified patterns in the geographic focus of the existing literature, supported by Table 6. Research tends to focus on different parts of the globe depending upon the topical sub-area. These trends often arise due to the availability of key RSD, or motivation to use REMOTE-ML due to limitations of energy data obtained through conventional (non-REMOTE-ML) methods. For example, energy access and reliability are especially relevant and challenging to measure in developing nations in Asia and Africa, and therefore the majority of studies in this sub-area have focused on those geographic regions. By contrast, Wind Resource estimation has focused more on European nations due to the relatively high wind energy penetration in these markets. Table 6 not only summarizes the geographic focus of each sub-area, but also reveals geographic gaps in existing research, representing opportunities for additional study. For example, Africa and South America are relatively understudied regions across most of the REMOTE-ML literature, and therefore represent opportunities to apply REMOTE-ML methods to novel regions. These geographic regions comprise many developing nations, which may be difficult to study using conventional data collection mechanisms.

*Opportunities and recommendations:* Broad-scale applications of REMOTE-ML to support national or global acquisition of energy data are currently limited but will be increasingly valuable. For these applications, the costs of commercial imagery are likely to be prohibitive. Impactful research areas for addressing this challenge include understanding which quantities of energy systems information are attainable at different spatial resolutions and identification of public remote sensing resources that can support REMOTE-ML objectives. This may require determining the lowest resolution of RSD for which the problem can be adequately addressed and using those data for upscaling to larger geographies. There are also opportunities to fill geographic gaps in the existing literature, as summarized in Table 6.

REMOTE-ML applications are also growing in related fields of environmental impacts and externalities associated with energy systems. The field of climate studies has been rapidly growing, and overlaps with this work; it explores many aspects of global emissions and impacts of energy systems, and has seen significant attention within the research community recently from monitoring power plant emissions plumes to agricultural monitoring and associated emissions using remote sensing data [252].

| Category | Asia | Africa | EU | NA | SA | Oceania | Other |
|---|---|---|---|---|---|---|---|
| Solar Potential Energy Estimation and Forecasting | [17], [27], [30], [32], [39]–[42], [49]–[52], [54], [55] | - | [20], [23]–[25], [34], [37], [43]–[45], [48], [53], [56] | [46] | [22] | [18], [26], [33], [38] | [31], [47] |
| Wind Resource Estimation and Forecasting | - | - | [62], [67], [72], [73] | [74] | - | - | [68]–[71] |
| Other Primary Resources: coal, geothermal, hydroelectric, and oil | [79], [82]–[89] | [81] | [78], [80] | [96] | - | - | [90]–[94] |
| Electricity Generation Infrastructure identification and characterization | [4], [109] | - | [108], [110] | [5], [97], [99], [101]–[103], [105]–[107], [253] | - | - | [105], [111] |
| Electricity Generation Infrastructure Risk Monitoring | [116]–[120], [122], [123], [125], [127], [129], [130] | [121] | [115], [124], [126] | [128] | - | - | - |
| Transmission and Distribution Line Mapping | [134], [139], [141], [143], [147], [150], [151], [154], [159]–[163] | - | [144], [146], [164] | [155], [156] | [167] | [136], [137] | - |
| Transmission Infrastructure Mapping & Monitoring | [176], [177] | - | - | [6], [172]–[174] | - | [169], [171] | - |
| High-voltage Insulator Monitoring | [162], [181], [183], [186], [194], [198], [200], [201], [206] | - | [197] | - | - | - | - |
| Building Energy Consumption | [211], [214], [216], [225] | [212] | [8], [223], [224] | [213], [222] | - | [208] | [215] |
| Electricity Access and Reliability Estimation | [219], [236]– [240] | [217], [232] | | [241] | - | - | [228], [235] |

*Table 6: Classification of review studies by subject and geographic area. NA: North America; SA: South America; EU: Europe (including UK)*

### 5.4. Machine learning challenges for remote sensing data

The following set of unique challenges may be faced by energy practitioners seeking to implement REMOTE-ML techniques.

*Domain adaptation.* Domain adaptation is the challenge of overcoming performance reductions when training a machine learning algorithm on data from one location (e.g. Tokyo, Japan) or sensor type (e.g. satellite RGB image) and applying it to a different location (e.g. Phoenix, Arizona) or sensor type (e.g. infrared data) [244]–[246], [254]. While a number of techniques have been proposed for these domain gaps (or distribution shifts), this remains an open question and a limitation on the ability to scale up or generalize the application of machine learning techniques [255] to larger remote sensing datasets.

*Rare object detection and the limitations of training data.* Identifying and monitoring energy system infrastructure is a goal of many studies in this review. However, this task presents the unique challenge that most energy infrastructure is relatively rare and occurs sparsely within imagery; for example, there are few power plants and transmission lines relative to the total area of a given country. With few examples to be found, this often makes gathering adequate training data more difficult and expensive, potentially limiting the types of analyses that can be conducted. In the ML literature zero-shot and few-shot learning research aims to develop models that can effectively learn from very few, or even zero, labeled training examples [256]. Other recent research has sought to circumvent this problem by creating large quantities of *synthetic* RSD that can be used to train ML models [244]. Despite significant research progress in these areas however, it remains challenging to create ML models for rare objects.

*Data fusion.* In many geographic locations there may be multiple information-rich sensor modalities, with each modality providing somewhat different information on the region that could be collectively used for REMOTE-ML. In such cases, it may be desirable to combine multiple forms of data (e.g. RGB imagery and synthetic aperture radar) and use all of it together - this is an open challenge of sensor fusion [257] that requires that we rethink modeling paradigms to accommodate the disparate sensor modalities into one learning framework.

*Opportunities and recommendations:* Few of these methods have been directly applied to energy system applications even though they present excellent opportunities in each of these areas. Each of the challenges in this section correspond to active areas of research for remote sensing including domain adaptation [246], rare object detection using zero- and few-shot learning or synthetic RSD [244], [256], and sensor fusion [258]. Integration of domain adaptation techniques could, for example, be applied to analyses that are currently constrained to regional studies to scale them up to the national or global scale.  Moreover, the detection of rare objects such as small-scale solar PV could be strengthened by utilizing recent zero-shot and few-shot techniques, or synthetic RSD.

## 5.5. Advances in remote sensing hardware, data collection, and availability

As satellite data providers have grown significantly over the past two decades, their offerings have become more diversified. Data providers have been trending towards distinguishing their products through increased resolution (e.g. Maxar), increased collection frequency (e.g. Planet), or greater accessibility (e.g. Landsat, Sentinel). Higher resolution enables greater visibility of smaller objects (like rooftop solar) and increased frequency enables more rapid assessment of changes (like power grid expansion in nations undergoing electrification efforts).

Simultaneously, numerous products have started to become available to ease the process of accessing these data through application programming interfaces (APIs), analysis platforms that draw public data (e.g. Google Earth Engine [259], Microsoft Planetary Computer [260]), and data ordering processes that are more streamlined and integrated into cloud-compatible workflows. Local and regional data collections are now also available with access to inexpensive drones that can capture custom, high-resolution photographs. Lastly, more types of sensor modalities beyond the collection of RGB data have become available with particular growth in infrared, synthetic aperture radar, and lidar sensors.

There are two major barriers in this space to greater use of remote sensing data: (1) cost of acquisition and (2) data management and processing. The most significant barrier is cost. Free, lower resolution offerings such as Landsat and Sentinel often lack the detail needed for energy system analyses with some exceptions for very large scale infrastructure or environmental impact monitoring. This typically means more expensive, proprietary datasets are required. For the highest resolution data, acquiring data for an area the size of a typical U.S. or E.U. state can easily run into the millions of dollars. The precise cost is highly variable, however, and will depend on imagery resolution, the quantity/scale of data, whether the data would require a new tasking of the satellite (a fresh collection), whether multiple images of the same region are required, and whether discounts may be available based on institutional status or for specific research objectives.

Once data are procured, preparing it for use and storing it presents additional challenges. While improvements are actively being made by many data providers, the process of acquiring proprietary RSD can be time-intensive.  This is especially true for those who are not trained remote sensing professionals or may be entering the space for the first time, and therefore may incur additional computational and storage costs in the process.

*Opportunities and recommendations:* While RSD resolution and collection frequency are improving, cost remains a major challenge to broader application of REMOTE-ML methods.  Continued reductions in the costs of proprietary data will be key to expanding their utility and uptake by the energy community. Secondarily, the accessibility of the data and analyses could be increased through the continued development of geospatial data processing platforms, like Google's Earth Engine. By enabling quick access to public datasets and APIs, they substantially lower the barrier to entry into this space.

### 5.6. Security, privacy, and ethical considerations

An often omitted, but nevertheless present and important set of considerations with each of these studies are the potential security, privacy, or ethical concerns that may arise. Security concerns are typically related to critical energy infrastructure data (such as oil and gas pipelines, large generators, substations, critical transmission arteries, etc. - anything needed to keep a country fueled and electrified). There are concerns and sometimes outright prohibitions around what type and level of detail of information can be provided [261] and these may vary regionally within a country as well as across borders. These laws are also rapidly evolving, for example UAV/drone regulations have been recently established, such as the Operations Over People rule from the U.S. Federal Aviation Administration, which went into effect in mid-2021.

Privacy concerns exist about Personally Identifiable Information that is made accessible through RSD. Concerns around violations of privacy are related to any information that may reveal individual behaviors or actions that otherwise would not be visible [262].

Lastly, there are other ethical considerations that come into play when the use of REMOTE-ML methods for assessing energy system infrastructure and planning for the future may intersect with issues of equity and environmental justice. For example, vulnerable populations have been largely underrepresented in historical data. Therefore, proximity to energy infrastructure that has the potential to result in negative environmental impacts (e.g. polluted rivers, water tables, etc.) for any historically marginalized community needs to be analyzed carefully to ensure that new development is responsive to vulnerable communities [263].

*Trends and limitations:* Data privacy and security is an area of active discussion as the rapid rise of data and computation has outpaced the complementary ethical and legal frameworks that guide this work [1]. As one example, in the U.S., programs ranging from the U.S. Department of Energy's DataGuard program for smart meter privacy [262] to the Federal Energy Regulatory's guidance and the continued policy development related to security around critical energy infrastructure information [264].

*Opportunities and recommendations:* Applications in the REMOTE-ML space have the opportunity to address these concerns as part of the research produced, and provide an assessment and/or justification for a study's design and in defining appropriate data-sharing protocols. In the end, these are ethical questions rather than strictly scientific ones, and care should be taken to consult institutional review boards or other ethics-focused resources if there is concern about privacy or security. A larger stakeholder-engaged process that includes civil society may be valuable in the longer term [265].

## 6. CONCLUSIONS AND FUTURE OUTLOOK

REMOTE-ML techniques are used in a growing body of work that brings together the strengths of RSD and ML tools for advancing energy system applications. Although REMOTE-ML methods are not necessarily capable of extracting all types of energy information, they have an important role to play in addressing gaps left by conventional methods of data collection. REMOTE-ML techniques are particularly useful and complementary to conventional energy data sources when energy system infrastructure data are not available, or when they are expensive to collect. Furthermore, these techniques can often be cost-effectively deployed for timely analyses, expediting the process of extracting critical insights required for efficient planning and operation of energy systems.

REMOTE-ML techniques have been shown to be promising tools for assessing, analyzing, and managing energy systems. The aspects of energy systems that have been investigated in the research literature span the energy value chain from primary resources (e.g. wind, solar, coal and oil) to end use consumption (e.g. building energy consumption estimation). While we identified REMOTE-ML applications related to conventional generation (e.g., oil/gas well, storage tank, and refinery identification) the vast majority of the studies focused on electricity systems. In electricity systems, REMOTE-ML techniques have been used for hazard identification (e.g., on generation infrastructure) by identifying possible warning signs on hydroelectric dams, wind turbine blades, and solar arrays using UAVs. Similar approaches have been used for transmission insulator monitoring. REMOTE-ML techniques have also been used to identify generation and transmission infrastructure including small-scale solar arrays and transmission line pylons/towers. REMOTE-ML techniques have even been applied to estimate electricity access and reliability of communities around the world using lights at night data.

While REMOTE-ML techniques offer many opportunities for new, and often more efficient, approaches to traditional energy systems analyses, we also identified a number of challenges that REMOTE-ML methodologies face for wider adoption and utilization in the energy sector. First and foremost, the quantification of data trustworthiness surrounding data accuracy, generalizability and bias is an area that is not often addressed in the existing REMOTE-ML literature. This has the potential to lead to overly-optimistic estimates of performance that could undermine confidence in models and derived insights that are not fully supported - rigorous performance evaluation offers the potential to avoid this pitfall. Similarly, few public benchmark datasets exist and varying validation techniques from study to study inhibit effective performance comparison across studies. The limited availability or publication of explanatory or validation data, often based on proprietary datasets, inhibits replication and incremental research innovation. Overcoming these deficits through the adoption of best practices within the community is a significant opportunity to increase innovation and practical applications within this field.

Lastly, ongoing developments in the fields of ML and remote sensing will also benefit REMOTE-ML applications. Research into the technical challenges in machine learning techniques such as domain adaptation, few-shot learning, and sensor fusion offer the possibility of increasing the performance of REMOTE-ML applications while reducing data requirements and cost. Parallel advances in remote sensing systems, for example, that decrease cost and increase resolution, will benefit accessibility and performance.

Supported by a large, rapidly growing and increasingly diversified literature on automated applications of remote sensing for energy system problems, REMOTE-ML techniques are poised to continue their evolution as a collection of high value tools for informing critical energy tasks.

Even as data sources become increasingly rich and ML techniques more powerful, attention to fundamental challenges and opportunities identified herein will be important to more completely deliver on the promise of automated energy systems data extraction, and which may be accelerated through establishment of research community practices that foster innovation and catalyze new frontiers. These techniques have the potential to provide the evidence needed for timely, reliable, and cost-effective energy systems planning, management, and policymaking.

7. Acknowledgements

This work was supported by NSF Grant Award No. OIA-1937137.

# APPENDIX

Detailed description of inclusion and exclusion criteria: We included a paper if all of the below criteria were met:

- Remote sensing (satellite, UAV, aerial like moving platform) data are used as input, excluding stationary remote sensing sources like mounted cameras, fixed position radars etc.

- Machine learning is (one of) the major method deployed in the paper, excluding papers that merely compare their result with a machine learning benchmark not implemented in their study

- The extracted information serves the energy community <u>directly</u>, excluding studies that focus on environmental effects like spills and leaks

| | Energy Terms | | Machine Learning Terms | | Remote Sensing Terms |
|---|---|---|---|---|---|
| 1 | Wind | 1 | machine learning | 1 | remote sensing |
| 2 | Solar | 2 | deep learning | 2 | satellite |
| 3 | Power system | 3 | support vector machine | 3 | aerial |
| 4 | Energy | 4 | random forest | 4 | UAV |
| 5 | Generator | 5 | regression tree | 5 | unmanned aerial vehicle |
| 6 | Coal | 6 | neural network | 6 | hyperspectral |
| 7 | Oil | | | | |
| 8 | Natural Gas | | | | |
| 9 | Geothermal | | | | |
| 10 | Hydropower | | | | |
| 11 | Power line | | | | |
| 12 | Transmission line | | | | |
| 13 | Electricity line | | | | |
| 14 | Energy infrastructure | | | | |
| 15 | Electric infrastructure | | | | |

the COMS MI Geostationary Satellite: A Case Study in South Korea," *Sensors*, vol. 19, no. 9, p. 2082, May 2019, doi: 10.3390/s19092082.